\documentclass[review]{elsarticle}

\usepackage{hyperref}
\usepackage{url}
\usepackage{booktabs}
\usepackage{subfigure}
\usepackage{graphicx}
\usepackage{float}
\usepackage[cmex10]{amsmath}
\usepackage{multirow}
\usepackage{longtable}

\begin{document}

\begin{frontmatter}

\title{Behavior variations and their implications for popularity promotions: From elites to mass in Weibo}
% \tnotetext[mytitlenote]{Fully documented templates are available in the elsarticle package on \href{http://www.ctan.org/tex-archive/macros/latex/contrib/elsarticle}{CTAN}.}n

% Group authors per affiliation:
\author{Bowen Shi, Ke Xu}
\address{State Key Lab of Software Development Environment, Beihang University, Beijing, China}
\author{Jichang Zhao$^*$}
\address{School of Economics and Management, Beihang University, Beijing, China\\
Beijing Advanced Innovation Center for Big Data and Brain Computing, Beijing, China\\
$^*$Corresponding author: jichang@buaa.edu.cn
}

\begin{abstract}

The boom in social media with regard to producing and consuming information simultaneously implies the crucial role of online user influence in determining content popularity. In particular, understanding behavior variations between the influential elites and the mass grassroots is an important issue in communication. However, how their behavior varies across user categories and content domains, and how these differences influence content popularity are rarely addressed. From a novel view of seven content-domains, a detailed picture of behavior variations among five user groups, from both views of elites and mass, is drawn in Weibo, one of the most popular Twitter-like services in China. Interestingly, elites post more diverse contents with video links while the mass possess retweeters of higher loyalty. According to these variations, user-oriented actions of enhancing content popularity are discussed and testified. The most surprising finding is that the diversity of contents do not always bring more retweets, and the mass and elites should promote content popularity by increasing their retweeter counts and loyalty, respectively. Our results for the first time demonstrate the possibility of highly individualized strategies of popularity promotions in social media, instead of a universal principle.

\end{abstract}

\begin{keyword}
\texttt Social media \sep User behavior \sep Elites \sep Mass \sep User influence \sep Popularity promotion
\end{keyword}

\end{frontmatter}

%\linenumbers

\section{Introduction}
\label{introduction}

Online social media, such as Twitter and its variant Weibo, are essentially reshaping the ecosystem of the conventional communication and thoroughly undermining the stereotypes of communication roles by replacing the mass media with new channels that are embedded in social networks \cite{brossard2013science,theocharis2016does}. Being both producers and consumers, instead of only audiences \cite{tsay2018mass}, massive users of social media are sharing and exchanging factual messages and mental statuses in a real-time manner, challenging the promotion of content popularity in essence. Content popularity is in fact the key goal for advertisers, innovators and influentials in communication, especially in marketing scenarios \cite{figueiredo2016trendlearner,lerman2010using,rogers2010diffusion}. Corporations have made efforts to find the right influentials for advertising and organizers even employ zealots to influence voters \cite{uzunouglu2014brand,mobilia2003does}. In line with this, word of mouth, the main form of information transfer in social media, heavily depends on the online influence and behaviors of its origins \cite{bickart2001internet,katz1955lazarsfeld,cha2010measuring,bakshy2011everyone,li2018attain}. The comparison of behavioral differences between influentials and mass grassroots is one of the important issues in communication \cite{zhang2016creates,wang2014big,su2019exploring,grossmann2018massCelite,beck2002elite}. However, the fine-grained behavioral variations and their implications for popularity promotions are rarely addressed. 
%Only a small group of people have influence \cite{rogers2010diffusion}, but both ordinary people and elites want to become famous or more famous to create trends \cite{zhang2016creates}. Consequently, the drivers of influence and popularity remain of interest to many researchers.

The Diffusion of Innovation theory has emphasized that influentials are the key elements in information cascades and are often referred as “opinion leaders”, “innovators” or “early adopters” \cite{rogers2010diffusion}. Given the ambiguity and lack of clarity of the phrase “opinion leader” \cite{boster2011identifying}, the term “elites” is employed to refer to influential users in this article, similar to many studies \cite{van2016information,chong2007theory}. In recent decades, many scholars focused on the comparison of behavioral differences between elites and the mass \cite{zhang2016creates,wang2014big,su2019exploring,gonzalez2013broadcasters}, but they ignore some key elements. On the one hand, the behavior variations across various user groups and content domains were ignored in previous efforts and still remain unclear. While being new channels of information exchange, in addition to average citizens, social media also provide diverse conduits for users such as news media, government agencies and enterprises \cite{barnett2017measuring}. Different user demographics might result in distinctive preferences and influence \cite{zhao2013inferring}. In their role as the government, Spanish authorities actively interact with citizens regarding local issues \cite{haro2018using} while enterprises only focus on promoting products or collecting online feedback \cite{karimi2015social}, accordingly their influence and popularity promotions should be treated differently. Moreover, Hilbert et al. summarized that communication contexts surely influence communication structures \cite{hilbert2017one}, and it is thus possible that various users may demonstrate different patterns in multiple domains. In specific, Usain Bolt definitely has a lot of fans in sport related domain in Twitter, while the mainstream media accounts such as BBC broadcast news on various aspects such as politics, society, sports and technology. Therefore, it is imperative that the behavioral differences between the mass and elites should be pictured across domains and user groups. On the other hand, the impact of behavioral differences between the mass and elites on their strategies to promote popularity has not been examined. Although a number of studies put forward some strategies to enhance content popularity \cite{figueiredo2016trendlearner,li2018attain,guan2014analyzing,wang2012active,ju2020new}, it is unclear whether these strategies are effective either for various user groups or various domains. In fact, how to enhance the popularity of posted contents and keep the passion of audiences has become one of the most important problems in social media marketing \cite{figueiredo2016trendlearner,lerman2010using}. In target marketing, marketers need to segment the market into homogeneous subsets to achieve maximum customer satisfaction \cite{langford2008strategic}. Therefore, for each elite and ordinary people who want to increase their influence, a mapping of strategy across user groups and domains is needed instead of simply imitating others. These reflections motivate this paper to explore the impact of behavioral differences on the content popularity, helping elites and mass with right actions of popularity enhancement in different communication contexts.

As a result of the scarcity of massive user data in social sciences and the complexity of multiple domains, many traditional methods (e.g., questionnaires and surveys) are challenging to implement because of the spatial limitations and high costs \cite{beck2002elite,rizwan2018using}. Fortunately, the digital traces accumulated and aggregated in social media provide a more efficient but less expensive proxy for investigating the exact mapping between user groups and content domains \cite{hu2017social,banos2013role,morone2015influence}. More importantly, social media is now an indispensable part of human life, and in 2017, the proportion of active users rose to a new high of 37\% worldwide, nearly 2.8 billion people \cite{zhang2018social}. Weibo has attracted 500 million users in China, surpassing any other social networking sites in China \cite{nip2016challenging}, and extensive efforts have been devoted to study user behaviors in Weibo \cite{zhang2016creates,wang2014big,guan2014analyzing}. In particular, the authentication category system of Weibo provides an opportunity to further study the fine-grained user categories of the mass and elites. Meanwhile, determining the appropriate number of domains is a difficult task because of the complex contents in Weibo, so we use a topic classifier suitable for the Weibo discourse system based on machine learning to divide the domains. In addition, considering that the status of elites should be constantly developing and changing in interaction \cite{berelson1968people}, users known as “Big Vs” but of no real influence will affect the results, and the traditional methods such as informants’ ratings and self-designation are subjective-biased and difficult to quantify the real influence of massive users \cite{rogers2010diffusion,xiong2018accumulation}. In contrary, we establish retweet networks to select elites that are really influential.

To investigate the comparison between the mass and elites across user groups and content domains in a data-driven manner, techniques and methods from machine learning and social network analysis are employed in this study. With the help of a topic classifier adapted to the discourse system of Weibo \cite{fan2015topic}, we use the machine learning model to divide 140,000,000 tweets into seven main topic categories such as society, sports and so on. Then, by collecting retweets of 8.52 million users in seven domains, seven networks are established to identify the elites. We apply the position of node in the topology to measure the importance of users \cite{momtaz2011identifying} and ultimately selected 930 truly influential users. As for the category of user authentication, unlike Twitter, Weibo has a strict verified system which requires users to provide manual documentary evidence and divides them into five main categories such as media, government and so on. In particular, these verified users play crucial roles in the information dissemination of Weibo. Accordingly, the verified types can be a direct clue for grouping users. In fact, grouping all users into different clusters, on the one hand, will support the investigation of all participants in online communication instead of only elites and, on the other hand, will greatly reduce randomness at the individual level and make it feasible to stably map user behavior onto content domains at the collective level. In terms of splitting contents into seven domains and apportioning users into these groups, the following investigation of how user behavior varies across content domains can be comprehensively conducted.

Tweeting and retweeting are the most frequent components of user behaviors in Weibo \cite{leal2014influence,wiertz2007beyond}, and Figueiredo et al. emphasized that other factors such as content links can also affect popularity \cite{figueiredo2016trendlearner}. On the basis of a well-established grouping system of users and domains, here the behavioral differences between the mass and elites are comprehensively probed from the perspectives of activity, homophily, loyalty, and content characteristics. We further attempt to figure out strategies fit for the behaviors of various users to specifically increase their content popularity. Through the comparison from multiple perspectives, many unexpected differences in behavior and strategy between the mass and elites are revealed. This study powerfully demonstrates that each user needs to choose the right ways to increase influence across domains, suggesting that the popularity promotion strategy is closely coupled with content domains and user groups. The exact mapping established here can directly help develop suitable strategies for popularity promotions in social media, which is particularly instrumental to market segmentation in target marketing \cite{langford2008strategic}. Taking the action of adding links as an example, we demonstrate that the mapping between user groups and content domains can inspire ways to enhance popularity in a fine-grained manner, especially as both the user group and the content domain are the inputs of this practice.  Additionally, the diverse perspectives are investigated, which further ensures the extendibility of our conclusions.

The main contributions of the paper are as follows:

(i) This study is the first to disclose the behavior variations from elites to the mass across user groups and multiple domains in social media. With regard to splitting users into five groups and contents into seven domains, an accurate and complete spectrum of behavior variations across domains is comprehensively established. With the help of spectrum, what kinds of users targeted as behaviorally influential seeds in marketing-like applications can be optimally pinpointed.

(ii) A comprehensive mapping between behavior variations and popularity promotions is established in rich perspectives ranging from activity patterns to various content characteristics. In particular, though targeting influentials is extensively exploited, this is the first time to study the popularity promotion for the mass. Appropriate strategies for popularity enhancement can accordingly be derived from the mapping in terms of taking both user groups and content domains into account.

(iii) Machine learning and network analysis are jointly employed in this study, which enriches the practical methodologies in probing massive users of communication study. Driven by massive tweets and huge retweet networks in Weibo, solutions involving artificial intelligence and intensive calculations are conducted to split user groups, cut content domains and draw the mapping, overcoming high costs and low efficiency of conventional approaches.

\section{Related works}
\label{sec:rworks}

\subsection{Differences in behavior between elites and the mass}
\label{subsec:bdiff}

In social media, everyone is simultaneously a publisher and a listener of information, and all users equivalently constitute the communicator and audience elements in the communication model \cite{tsay2018mass}. According to the two-step flow model \cite{lazarsfeld1944people}, the propagation of information is a secondary dissemination process in which most people form their own views under the influence of elites, e.g., public opinion leaders. Opinion leaders characterized as the influentials with more connections are crucial for information dissemination \cite{luarn2014network,weimann1994influentials}. In the meantime, the influentials hypothesis, in which influentials will trigger wide dissemination, has been questioned in recent years \cite{zhang2016creates,watts2007influentials}. It has already been pointed out that the mass play a decisive role at the early stage of trend creation \cite{zhang2016creates}, implying that user influence can be counterintuitive and cannot be overly simplified, and elites and the mass have gap in opinion \cite{grossmann2018massCelite,beck2002elite}. Therefore, the comparison between elites and the public is a meaningful task that deserves more efforts.

As a key issue, many studies regarding the comparison between elites and the masses inherently neglected the behavior variations across user groups and content domains \cite{zhang2016creates,wang2014big,su2019exploring}. In particular, various user demographics, e.g., professions, might result in different behaviors in online social media \cite{zhao2013inferring}. Enterprises collect online feedback \cite{karimi2015social} while athletic stars promote peacebuilding activities in Kenya \cite{wilson2015celebrity}, and Zhao et al. divided users into four categories (i.e., engineer, recruiter, salesperson) that fit for Linkedin to study their behavioral differences \cite{zhao2013inferring}. Moreover, Smith et al. observed six different communication patterns in digital media \cite{smith2014mapping} and contexts were also emphasized to feature different structures \cite{hilbert2017one,guisinger2017mapping}. In understanding the behavior variation over multiple domains, it is also possible that elites may demonstrate patterns that differ from others. In this paper, to capture a complete picture, the behavior variation of elites across content domains and user groups is therefore separately discussed and compared with the case of all users, i.e., the mass level.

\subsection{Behavior for popularity promotions}
\label{subec:bpp}

User behavior is a direct reflection of the information diffusion, in which tweeting and retweeting are two primary activities in Weibo and have been exploited extensively in previous efforts. As the frequency of posting, high activity indicates a greater likelihood of exposing \cite{gonzalez2013broadcasters,wang2012active,leal2014influence}. Gao et al. suggested that tweeting may vary differently in working hours and leisure time \cite{gao2012a}. Many users like to embed links of images, videos and news to make contents charming in social media \cite{casalo2018influencers}. While retweeting is a crucial attribute in interactive behavior \cite{wiertz2007beyond,boyd2010tweet} and reflects the social homophily. The homophily refers to the fact that individual prefers to contact with the people with many similar behavioral characteristics \cite{mcpherson2001birds}, and has been demonstrated in various social media \cite{kwak2010twitter,colleoni2014echo,zhou2018homophily,centola2010spread}. Loyalty, another factor that impacts retweeting, is measured by the retweeting frequency and reflects the multiple behavior properties in essence such as interaction, cooperation, and intimacy. Nevertheless, the comparison between mass and elites on behavior is rarely performed on these different dimensions, implying the necessity of more comprehensive explorations. 

Content popularity is the prime target in communication. Many factors underlying behaviors can affect the content popularity, in particular the narrative characteristics~\cite{figueiredo2016trendlearner,rogers2010diffusion}. Intuitively, rich and diverse contents will attract more audiences of different interests, but cognitive psychologists have long contended that human beings have a limited capacity for information processing \cite{zhu1992issue}. Too many kinds of topics may lead to a decrease in content quality in a single domain, thereby losing audiences and even popularity. It is also indicated that users in Weibo are quite keen on inserting short links jumping to news, pictures and videos into tweets~\cite{wang2012active,guan2014analyzing,yu2011beyond}. In addition, the loyal customers play an important role in maintaining a basic level of attentions \cite{shin2004segmentation}, and increasing loyalty can upgrade profits \cite{cheng2009classifying,boateng2019online}. These factors could be potential features in popularity prediction and promotion. Szabo and Huberman used linear regression to predict the online popularity in YouTube and Digg \cite{szabo2010predicting}, while Chen et al. applied a binary classification model to identify the trend in time series \cite{chen2013a}. However, they ignored the behavioral differences of various user clusters across content domains, after all, Figueiredo et al. highlighted that the domain’s context is an crucial factor in changing popularity \cite{figueiredo2016trendlearner}. Meanwhile, there is a lack of fine-grained recommendation on effective enhancement strategies. In this paper, content domains, user groups and other dimensions of behaviors such as loyalty, content diversity will be comprehensively integrated to target the right enhancement strategy of each situation.

\section{Data and methods}
\label{sec:dandm}

\subsection{Weibo data set}
\label{subsec:dataset}

The Weibo data in this study were collected through its open API (application program interface). Over 140,000,000 tweets from the Weibo stream occurring from 10 October 2016 and 10 January 2017 were continuously crawled and in total, we sampled 8,520,933 unique users. The signals delivered in these posts are sophisticated and are from every aspect of everyday life. Specifically, the JSON file of each tweet contains attributes of text, retweet status and user demographics such as the verified type, gender, address and the number of followers, suggesting that the content domains, user groupings and influence metrics can comprehensively be derived from these attributes. For each user, the tweeting frequency and retweet times are accumulatively counted based on the retweet status of the user’s tweets, and the other rarely updated demographics, such as gender and verified type, are obtained from the latest tweets in our data set.

\subsection{User groups}
\label{subsec:user}

In particular, unlike Twitter, a distinctive verification mechanism in Weibo ensures the reliability of the user demographics, especially the verified types. In Weibo, users with certain verified types are known as the “Big Vs” \cite{wang2014big} and the platform even demonstrates red or blue badges on their profiles. Specifically, in addition to the basic real-name certification for each ordinary user, further verification steps involve (1) a certain reputation and influence in specific domains, (2) well-known enterprises and their executives, (3) the mainstream media, and (4) government agencies such as public authorities. Note that verification requires documentary evidence and is manually performed. More rigorously, enterprise users need to complete an Enterprise User Certification Information Form and Corporate Certification Application Letter and affix their corporate color seal and pay an annual fee. In general, the official verified types can be categorized in terms of the media, celebrity, government and enterprise. According to verified types, we can split the users into five groups, with the addition of those without verified types, i.e., ordinary users. Note that the authenticity of ordinary users can also be ensured due to the real-name certification regulation in China. The summary statistics of the user groups are in Table 1, with ordinary users accounting for the most and the government accounting for the least.

%\section*{Tables}
%\begin{table}[h!]

% \usepackage{multirow}

\subsection{Domain classifier}
\label{subsec:domain}

The main form of content in Weibo is text, and its topic can well represent the domain the content belongs to. Considering the massive text data, an automatic topic classifier is expected, and what’s more, the appropriate number of domains is critical. In this study, a previously well-developed Naive Bayesian classifier is adopted to perform domain categorization \cite{fan2015topic}. The classifier is trained on more than 410,000 Weibo tweets and its seven topic categories fit well with the news taxonomy of Weibo. Based purely on text features, the domain classifier can divide a tweet into one of seven topics: society, international, sports, technology, entertainment, finance and military. 

Both the F-measure and accuracy of the classifier in the cross-validation experiment is more than 84\%, suggesting its sufficient competence in the domain classification task. Concretely, we can first convert the text of each tweet into a vector {$\it{w_i}$}, where $\it{w_i}$ and $\it{i}$ refer to a term and its position in tweet $\it{t}$ after the word segmentation. In the incremental training process, the prior probability of term $\it{w_i}$ belonging to topic $\it{c}$ is calculated as 

\begin{equation}
P(w_i||c)=\frac{n^c(w_i+1)}{\sum_{q}{n^c(w_q)+1}}, \label{eq:1}
\end{equation}
where $\it{c}$ belongs to the topic categories $\it{C=(c_1, c_2, c_3, c_4, c_5, c_6, c_7)}$ and $\it{n^c(w_i)}$ indicates the count of occurrences of $\it{w_i}$ in topic $\it{c}$. Finally, the domain of a word vector is obtained by the maximum value of the probability calculated as $\it{P(c|t)=arg max_c P(c)P(w_i||c)}$, where $\it{P(c)}$ is the prior probability of $\it{c}$. Note that tweets labeled “unknown” by the classifier will be omitted in our analysis, due to the lack of confidence in determining their domains. The average precision of this classifier is convincing, and in particular, the large number of tweets that we employ in the experiment can further guarantee its accuracy after the aggregation. Its mechanism of incremental training can also solve the problem of new words in to-do tasks. In terms of grouping users into five clusters by user groups, angles from both user groups and content domains can be thus established to investigate behavior variations.

\subsection{Selection of elites}
\label{subsec:elites}

\begin{figure}[!hb]
\centering
\includegraphics[width=0.9\textwidth,trim=25 130 25 120,clip]{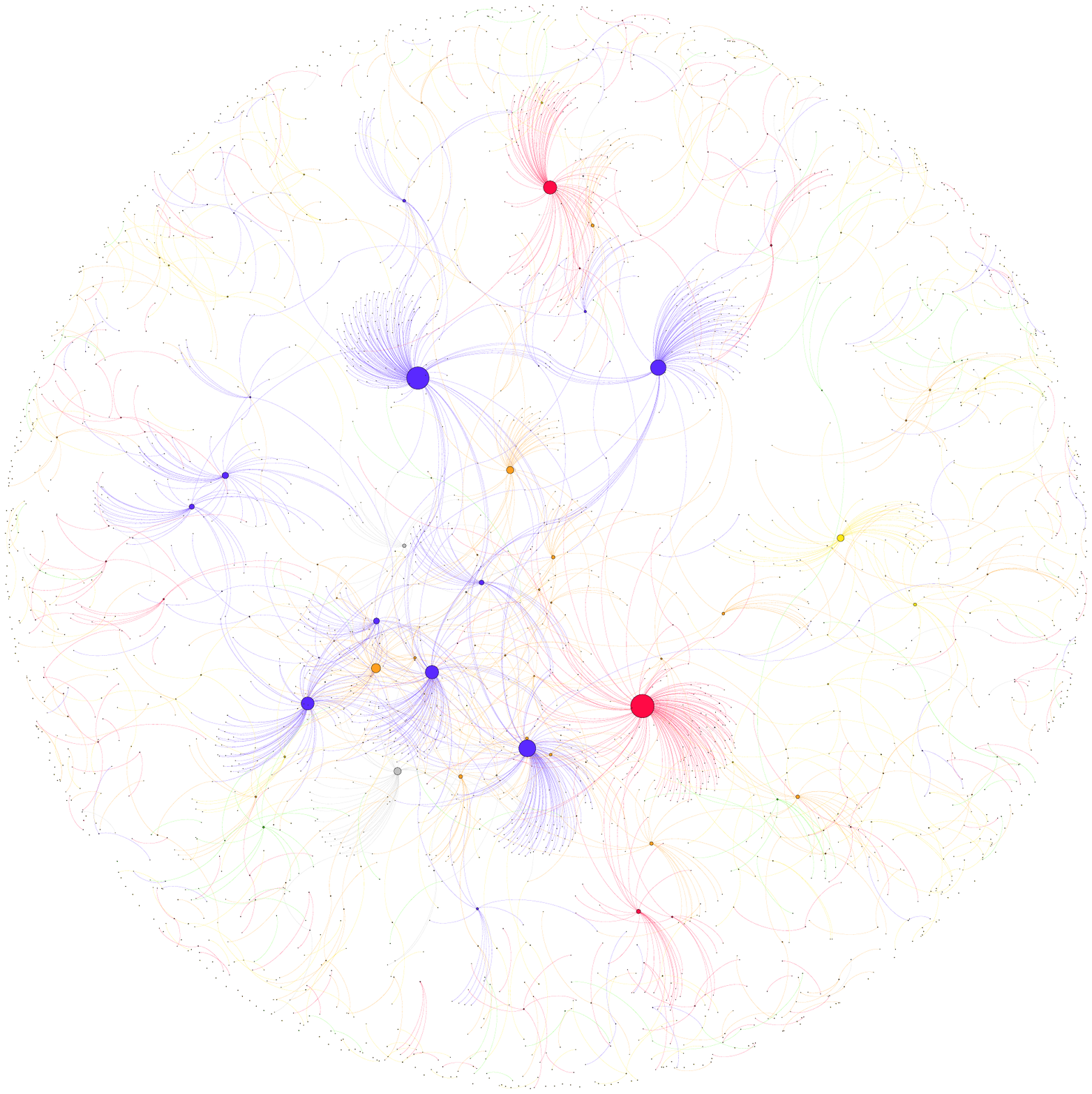}
\caption{{\bfseries Retweet network of various users in the military domain.}
      The threshold of the edge weight is set to 10, and the size of the node is related to its number of retweeters. We color each node by its verified type, i.e., blue represents the media, green represents enterprises, red represents the government, orange represents celebrities and gray represents others. Note that the color of the edge is the same as that of the source node.}
\label{fig:fig1}
\end{figure}

The formation and development of elites is a dynamic process, this status is constantly changing by quantifying the interactive behaviors \cite{lazarsfeld1944people}. Many researches' methods for selecting elites are too simple and rely on official verification \cite{zhang2016creates,wang2014big,su2019exploring}, and some users with “Big-V” may not be influential. User influence is essentially a reflection of interaction capabilities and therefore this paper targets the real elites through a lens of interactive networks. Weibo features a variety of interactive forms such as following, mentioning and retweeting. Needless to say, the frequency of being forwarded, through which tweets are disseminated in social media, is relatively more realistic and direct than the number of followers in reflecting user influence \cite{bliss2012twitter}. Moreover, the attributes in the Weibo data collected contain the retweeted status of original tweets and the corresponding author information; accordingly a retweet network between users can be constructed by extracting their retweeting relationships. The retweet network can be represented by a directed weighted graph, in which the nodes represent Weibo users (those without edges are omitted), the edges are the set of retweet relationships among users, and the weight of the edge is the total number of occurrences of retweets between user pairs (in our sampling period). The larger the edge weight is, the more faithful the retweeter is to the original publisher. Accordingly, we built seven networks using the separate retweet data from the seven domains in our later explorations. Fig.~\ref{fig:fig1} shows a sampled snapshot of the military retweet network with an edge threshold larger than 10 retweets for better visualization. These constructed retweet networks provide decent preconditions for subsequent work, such as the selection of elites and the inference of the user influence indicators.

\begin{figure}[] 
\centering
\subfigure[Mass]{\includegraphics[width=11cm]{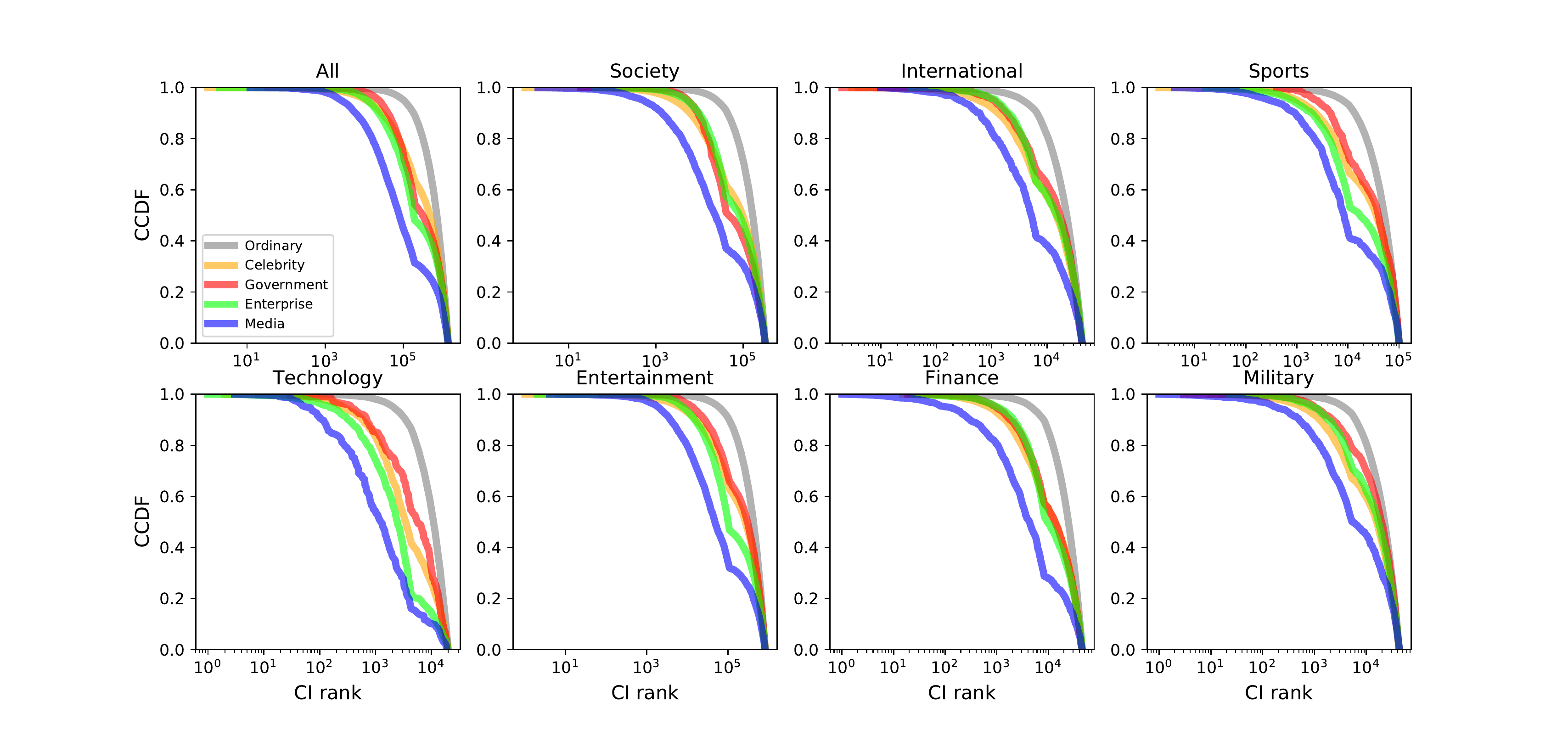}}
\quad
\subfigure[Elite]{\includegraphics[width=11cm]{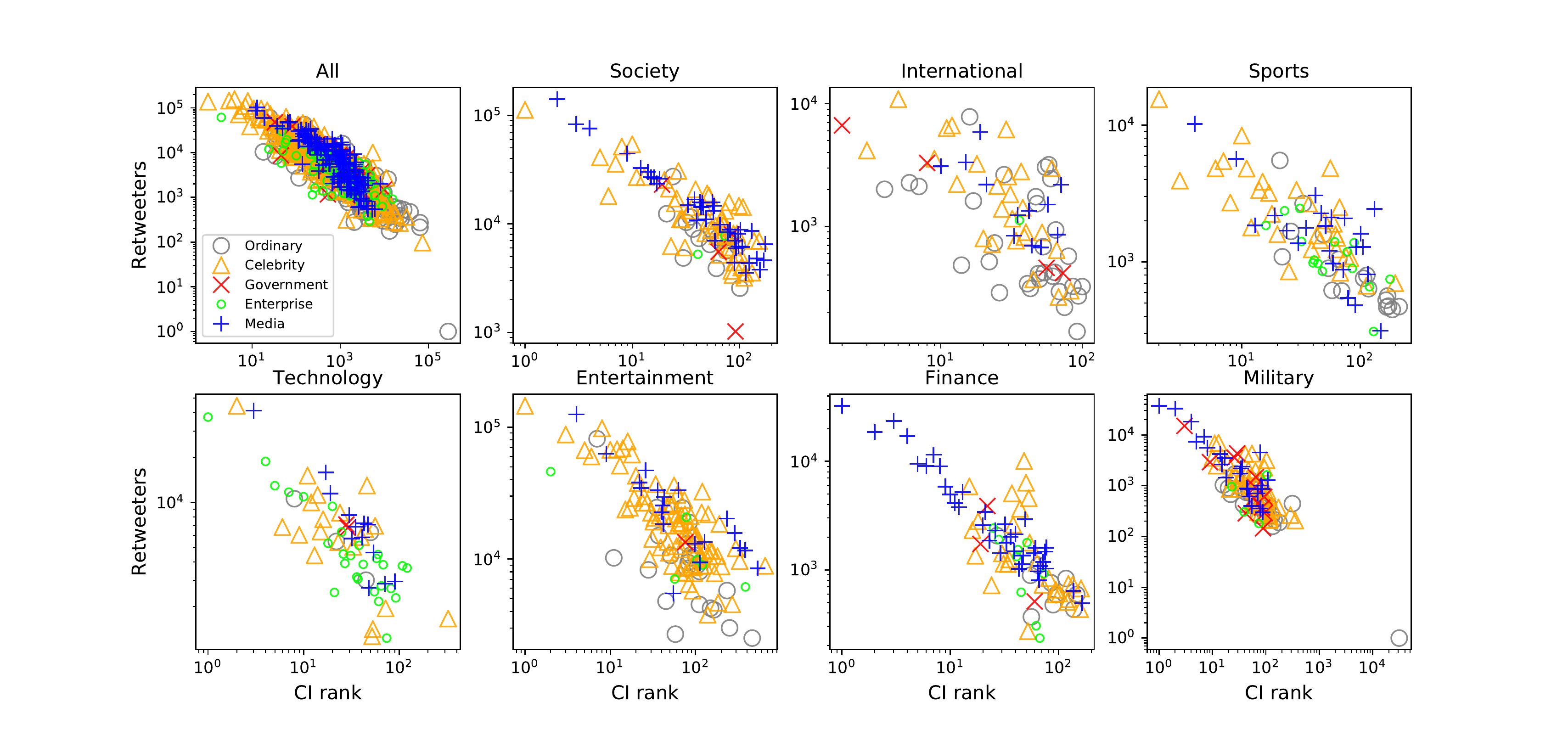}}

\caption{{\bfseries Mapping of user influence in retweet network and content domains.} Note that here, the CI is calculated within three hops as recommended and the sub-graphs (a) and (b) represent the mass level and the elite level, respectively.} 
\label{fig:fig2}
\end{figure}

The key element in marketing and information diffusion is a minority of influentials \cite{rogers2010diffusion}. More interestingly, after building a network through the retweet relationships between users, it is important to acknowledge that there can be many structural indicators for valuing user influence, such as in-degree, closeness, betweenness, many random walk methods and CI (Collective index) \cite{morone2015influence,nguyen2013social,ahajjam2018identification}. Al-garadi et al. mentioned that centrality methods have high computational complexity \cite{al2017identification}. For these seven large-scale networks in which the weight of each edge more than 2, the CI and the in-degree are employed to rank users by influence in each domain, and their computational complexity is O($\it{NlogN}$) and O(1) (where $\it{N}$ is the number of users in the retweet network), respectively. Due to the uneven size distribution of user groups, we select the top $\it{k}$ users as elites from all users, where $\it{k}$ is set to 200 and later experiments on a wide range of values also confirm the insensitivity of the results to $\it{k}$. The CCDF (complementary cumulative distribution function) and scatter plots of the CI rankings for the mass and elites are demonstrated across domains in Fig.~\ref{fig:fig2}. Note that the lower value of CI ranking represents more influence, and Fig.~\ref{fig:fig2}(b) shows that both indicators are positively related and elites selected from them are almost the same. According to in-degree, the distribution of elites in each group as $\it{k}$=200 can be found in Table 1, and a total of 930 unique elites are obtained from all domains. In addition, the influence changes of elites are more diversified, which is significantly different from that of the mass. For instances, enterprises dominate in technology, and celebrities are even more influential than media users in sports and entertainment. These differences between the mass and elites imply that user groups and content domains should be comprehensively considered, and the following experiments on behavior variations will be profiled and demonstrated at both levels of elites and mass.

\section{Behavior variations between the mass and elites}
\label{sec:bvaraitions}

After splitting users into five groups and contents into seven domains, how user behavior varies from the mass to elites can then be fully investigated. Focusing on the two primary behaviors of tweeting and retweeting, behavior variations will be specifically examined from views of tweeting activity, temporal patterns, homophily, loyalty and content characteristics, which together reconstruct a full angle of individual behavior in Weibo.

\subsection{Tweeting}
\label{subsec:tweeting}
\subsubsection{Tweeting activity}
\label{ssubsec:ta}

\begin{figure}[h!] 
\centering
\subfigure[Mass]{\includegraphics[width=11cm]{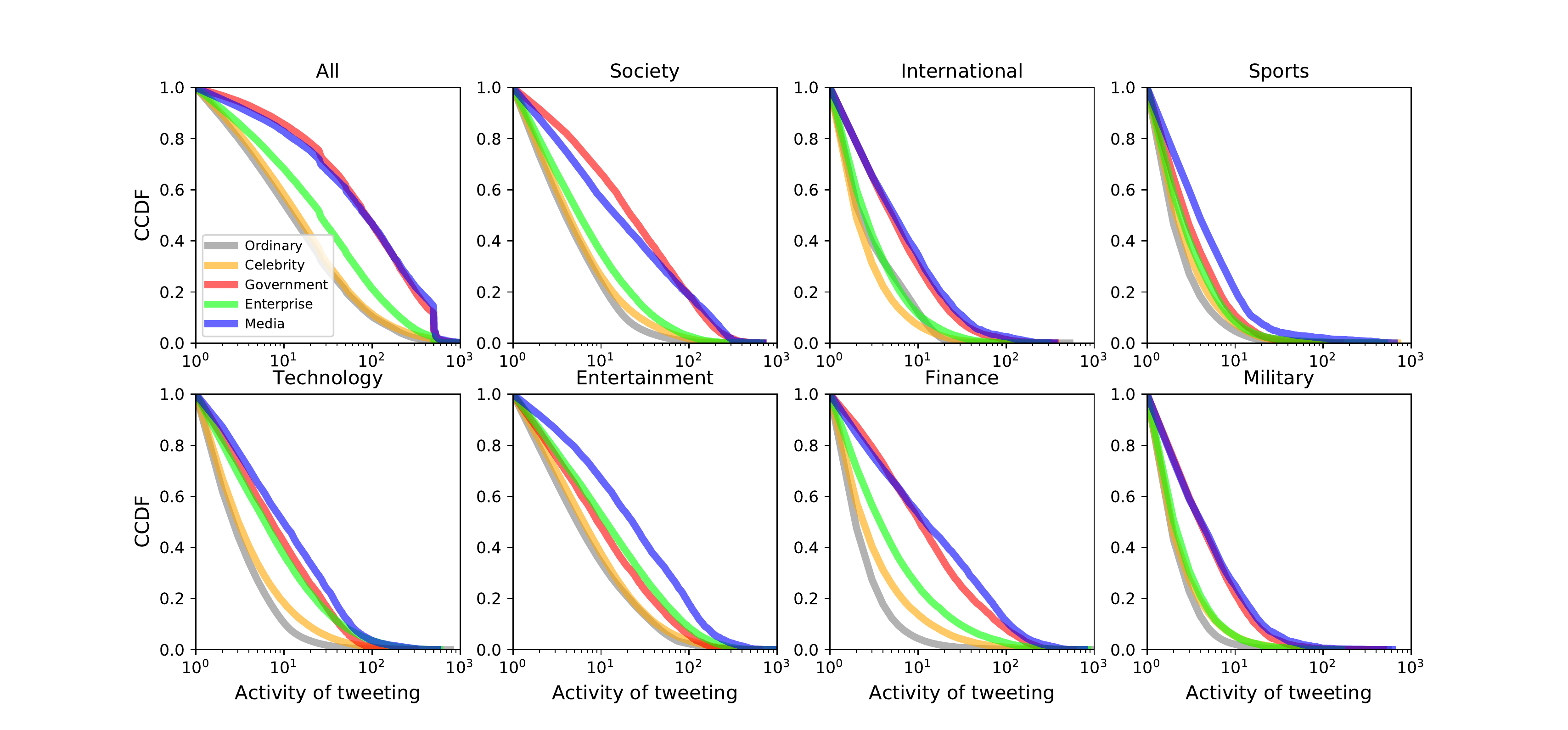}}
\quad
\subfigure[Elite]{\includegraphics[width=11cm]{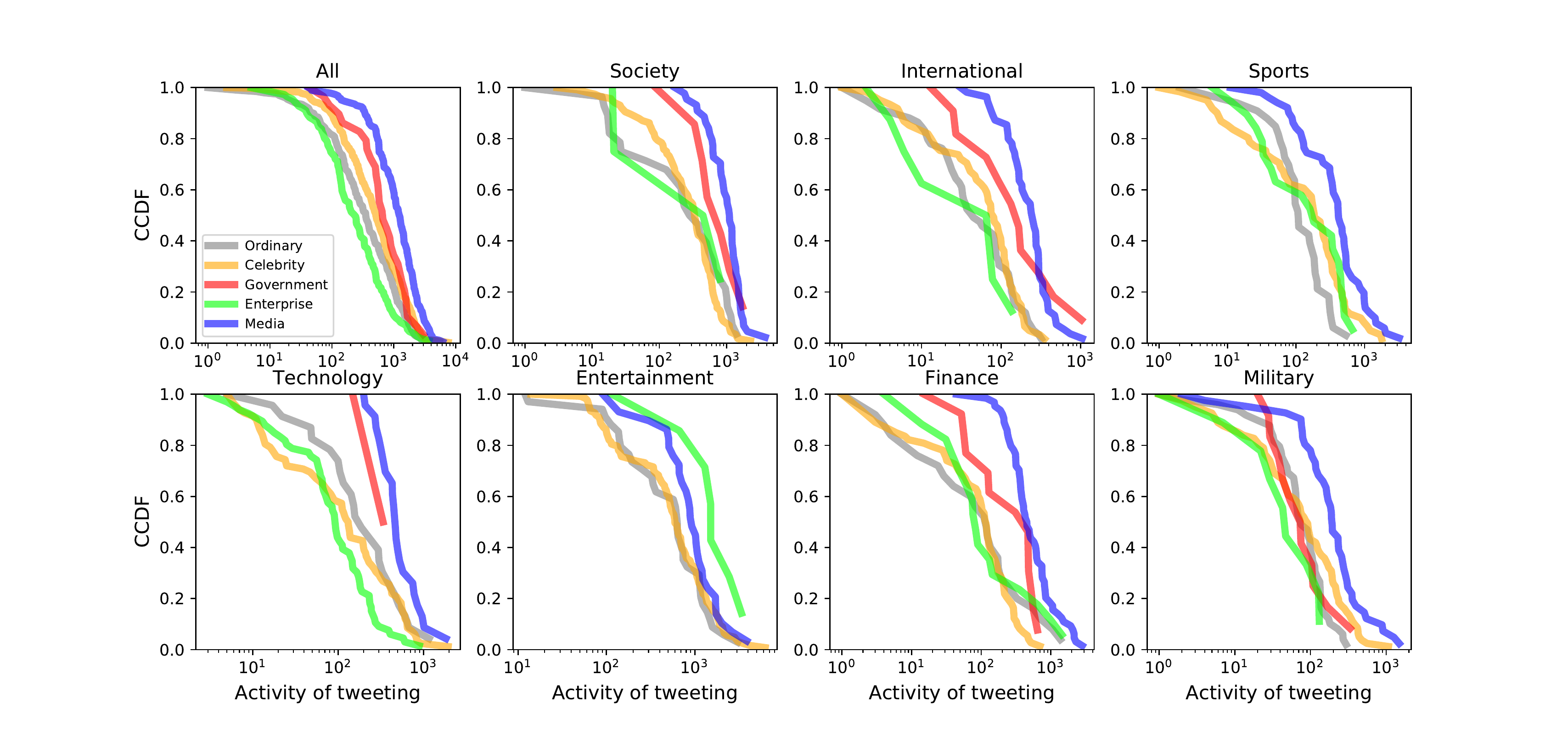}}

\caption{{\bfseries Mapping between tweeting activity and content domains.} Note that sub-graphs (a) and (b) represent the mass level and the elite level, respectively.} 
\label{fig:fig3}
\end{figure}

Posting more tweets, i.e., being more active in social media, will bring more opportunities to be noticed \cite{gonzalez2013broadcasters,wang2012active,leal2014influence}. Here, the activity of tweeting is simply measured by the number of tweets within the sampling period. For each user group, we obtain the CCDF (complementary cumulative distribution function) of activity in the seven domains at both the mass level and the elite level in Fig.~\ref{fig:fig3}.

At the mass level, for all domains, as shown in Fig.~\ref{fig:fig3}(a), the media has the highest proportion of active users in almost every domain except society and finance. It is counterintuitive that the activity of celebrities is relatively low and even lower than that of the ordinary users in the international domain. However, at the level of elites, the activity of various users has different patterns across domains, which is different from the situation of the mass. Surprisingly, the government elites even vanish in the sports and entertainment domains. In general, the elites have a higher level of activity than the mass and their patterns of varying across domains are also different. 

\subsubsection{Temporal activity}
\label{ssubsec:tempa}

\begin{figure}[h!] 
\centering
\subfigure[Mass]{\includegraphics[width=11cm]{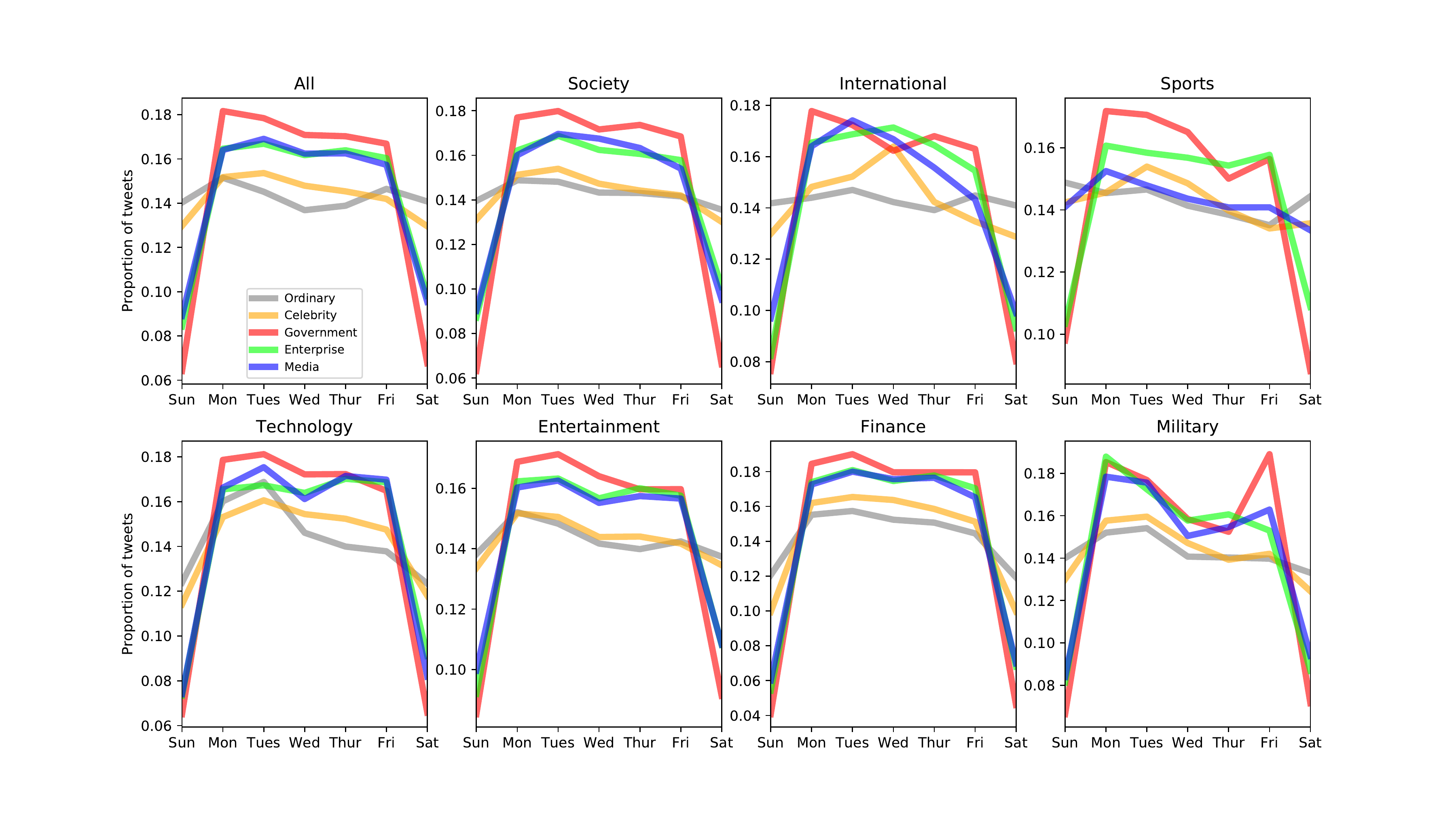}}
\quad
\subfigure[Elite]{\includegraphics[width=11cm]{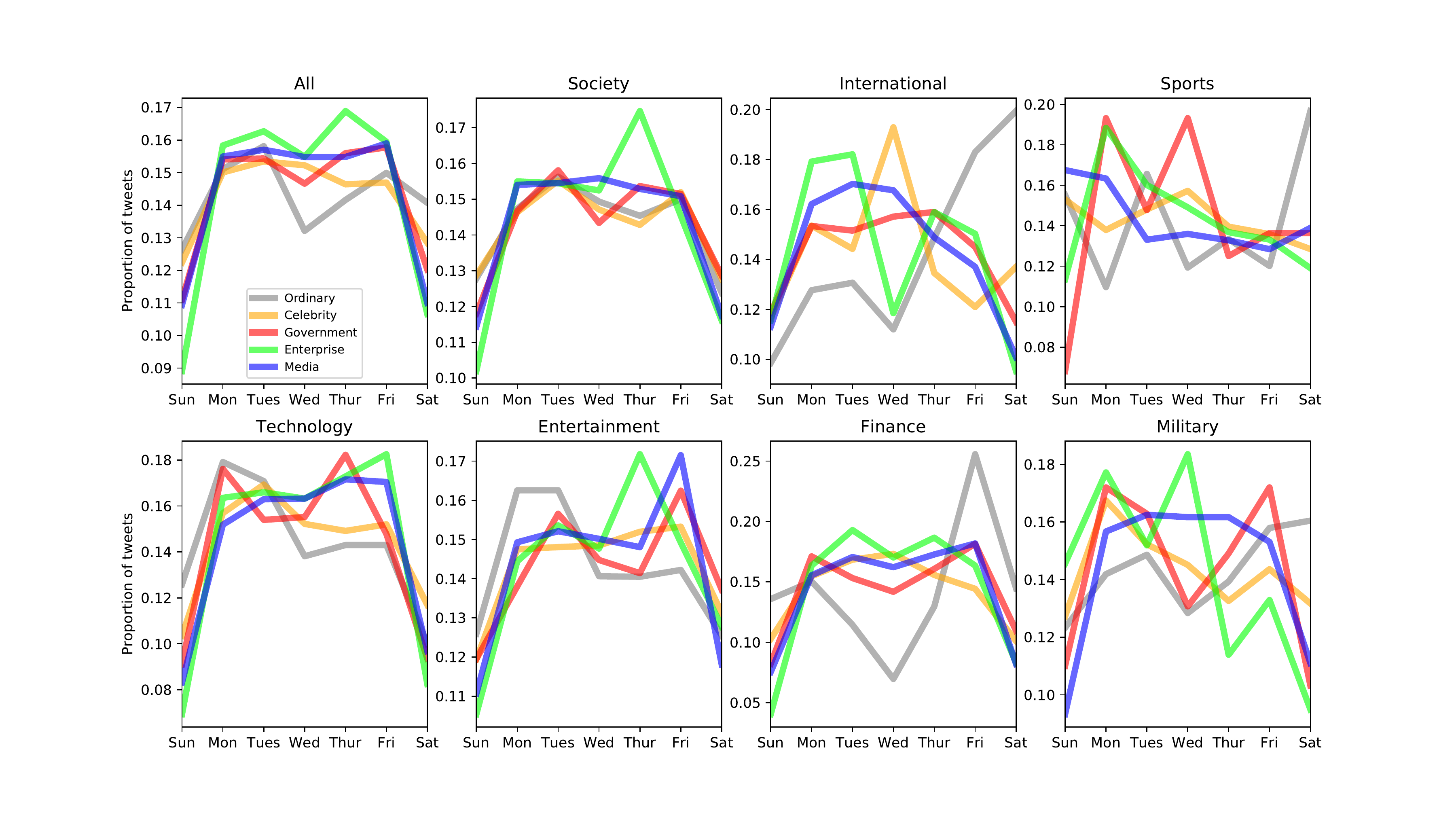}}

\caption{{\bfseries Temporal variation of tweets proportions across user groups and domains in a week.} Note that sub-graphs (a) and (b) represent the mass level and the elite level, respectively.} 
\label{fig:fig4}
\end{figure}

In this view our goal is to find out temporal patterns like weekly or hourly of various users across domains. Fig.~\ref{fig:fig4} shows the percentage fluctuations of tweets posted in a weekly manner and the ANOVA (analysis of variance) of the differences in weekly activity between the mass and elites across user groups and time periods is shown in Table 2. On the whole, the differences in temporal activity between the mass and elites are statistically significant, especially for the government and media groups. The elites of the ordinary and the celebrity are more active on weekends than mass, while other user groups have more obvious working cycles which can be explained that they might hire professional staffs to manage their accounts. There are exceptions for the media in the domain of sports, which can be explained that many games such as Serie A and La Liga are organized on weekends. The curves of various elites fluctuate sharply across domains, implying the diversity of elites' activity. In addition, the “double-peaks” curves in daytime indicate that government, enterprise and media users also possess working patterns in a day (see Appendix Fig.A1).

\subsubsection{Content characteristics}
\label{ssec:cc}

\subsubsection*{Content diversity}

\begin{figure}[h!]
  
      %\includegraphics[width=10cm,trim=50 0 50 80]{pic/f2a.pdf} 
      %\includegraphics[width=10cm,trim=50 80 50 80]{pic/f2b.pdf} 
%\subfigure[Mass]{\includegraphics[width=11cm]{pic/f2a.pdf}}
\includegraphics[height=8cm, width=12cm]{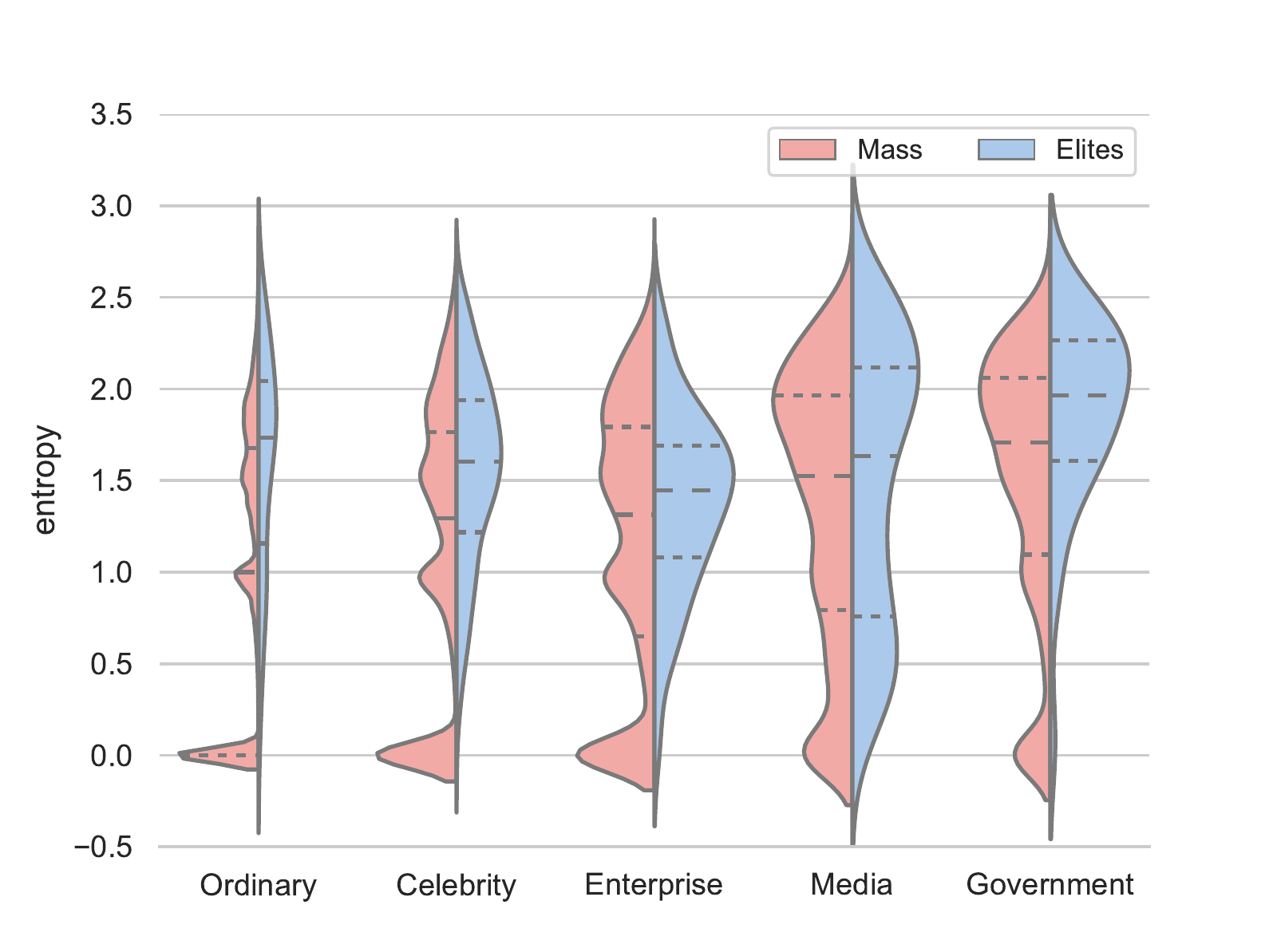}  %height=4.5cm, width=7.5cm
%\quad
%\subfigure[Elite]{\includegraphics[width=11cm]{pic/f2b.pdf}}

      \caption{{\bfseries The distribution of posting entropy for the mass and elites across user groups and domains.} }
      \label{fig:fig5}
      \end{figure}

To measure the diversity of content posted by various users, we calculate the posting entropy $H$, i.e., $$H=-\sum^{6}_{i=1}{p_ilog(p_i)}$$ of different topics, where $p_i$ refers to the proportion of posted tweets in domain $i$. The distribution of posting entropy of various users is shown in Fig.~\ref{fig:fig5}. To begin with, a certain percentage of mass users demonstrate a single interest, i.e., their posts only related to one domain and the posting entropy correspondingly equals to 0. Contrarily, elites post contents of richer topics than the mass, and their average value of entropy is accordingly higher than that of mass. Except for the enterprise, the top quartile of elites in other verified types is larger than that of mass, which is explained that the contents posted by elites in the group of enterprise are relatively unitary. 

\subsubsection*{Content links}

Users in Weibo would like to publish contents containing short URLs (t.cn) jumping to images, news and videos to attract audiences \cite{wang2012active}. To perform the analysis of content links, we transform the short links to the corresponding source URLs through Python package urllib2. Due to the speed limit with regard to tracing the source addresses of short URLs in Weibo, in this study, 100,000 users at the mass level were randomly selected to compare with the elites.

\begin{figure}[h!] 
\centering
\subfigure[Mass]{\includegraphics[width=11cm]{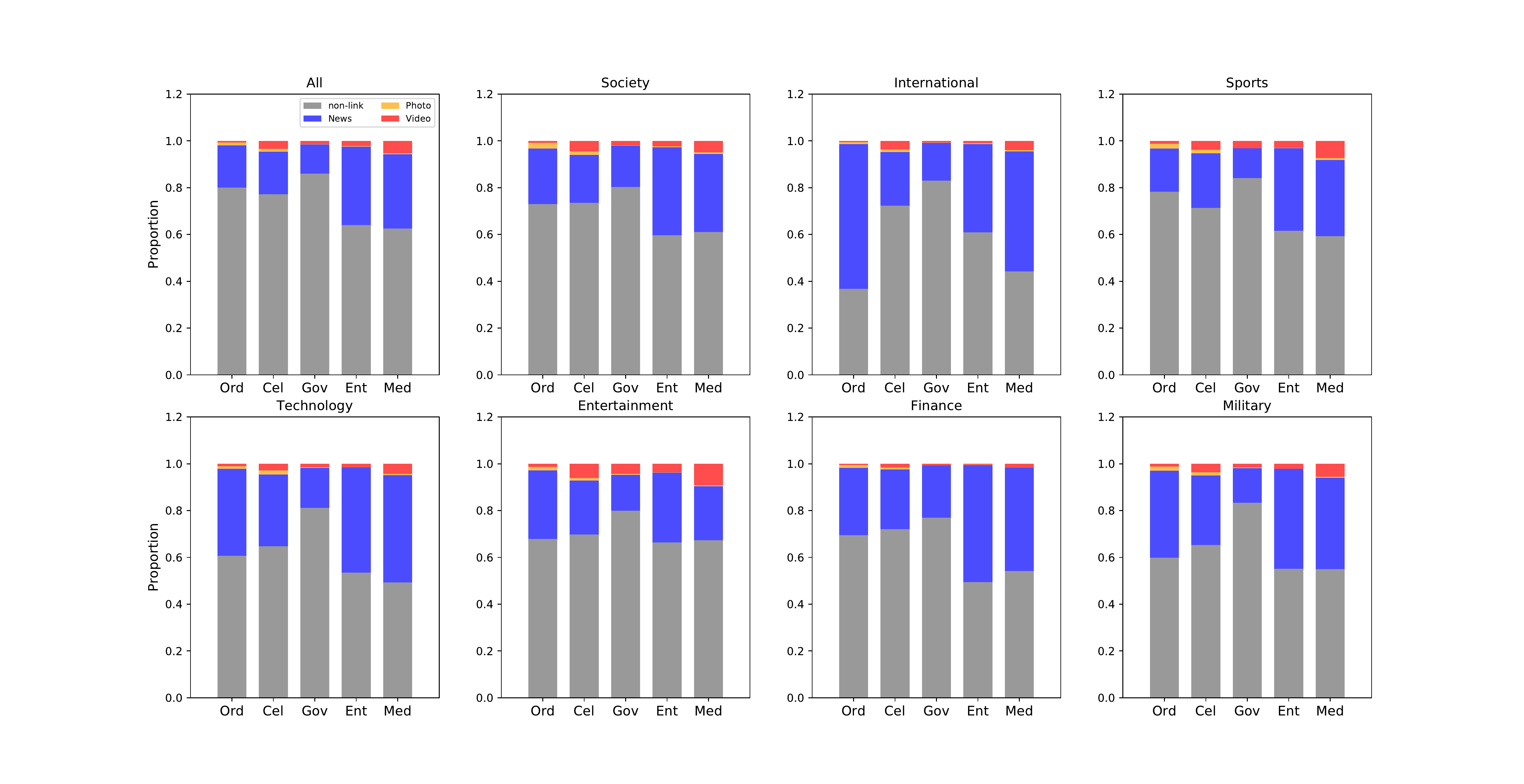}}
\quad
\subfigure[Elite]{\includegraphics[width=11cm]{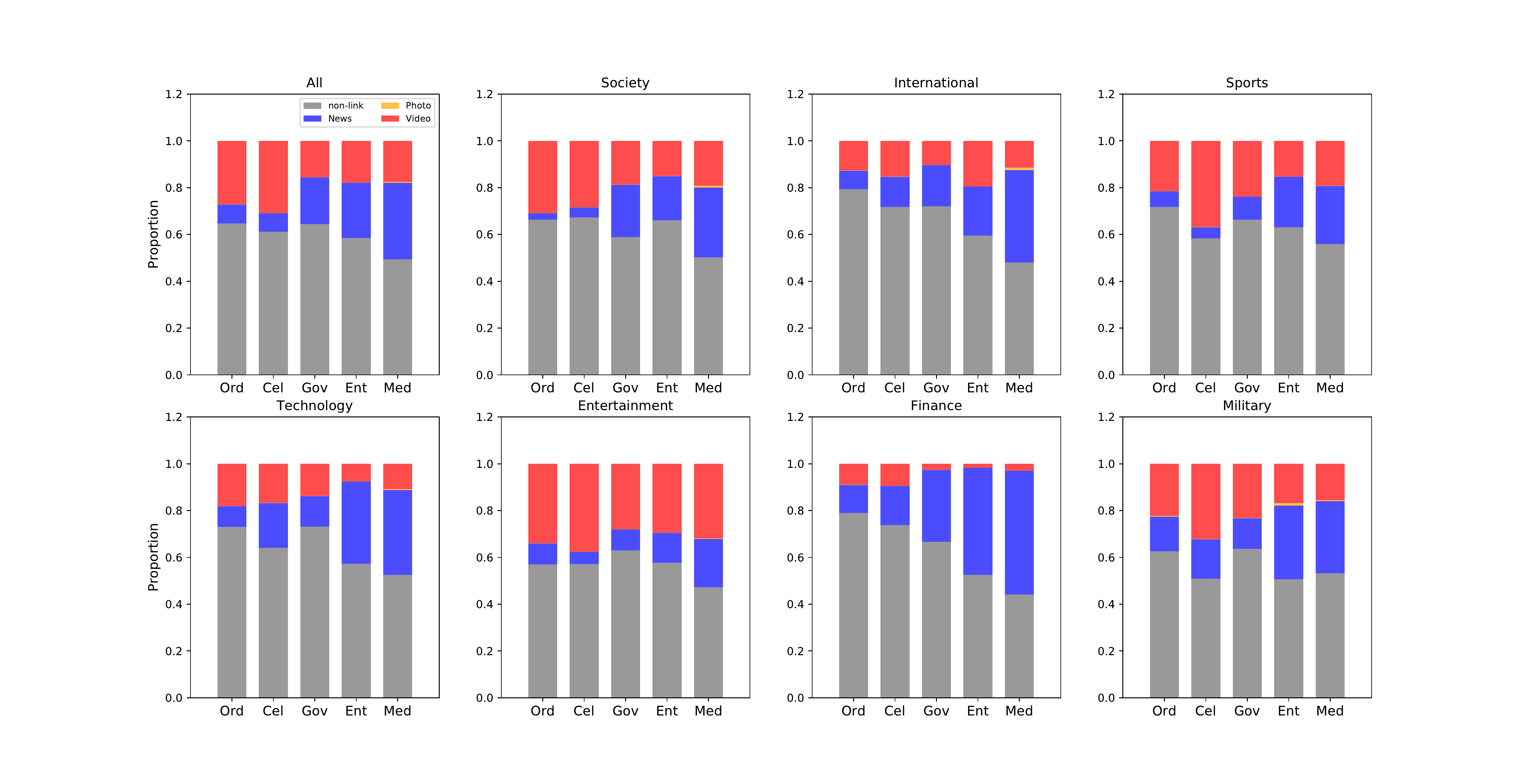}}

\caption{{\bfseries Percentages of tweets containing various links across user groups and domains.} Note that sub-graphs (a) and (b) represent the mass level and the elite level, respectively.} 
\label{fig:fig6}
\end{figure}

The percentages of tweets containing links at the mass and elite levels are respectively shown in Fig.~\ref{fig:fig6}, which illustrates the differences on content links across domains and user groups. In general, the elites obviously prefer to post tweets with video links, especially the celebrities. And the media has the largest proportion of using links no matter at mass or elite level. On the contrary, contents of the government are more formal, usually only words, which is in line with the previous finding that many accounts just posts government documents and may lose audiences \cite{tsay2018mass}. These differences may result in different content popularities, and the correlations between these posting preferences and retweets obtained will be further examined in actions of popularity promotion.

\subsection{Retweeting}
\label{subsec:retweeting}

\subsubsection{Homophily}
\label{sssec:homo}

\begin{figure}[h!] 
\centering
\subfigure[Mass]{\includegraphics[width=11cm]{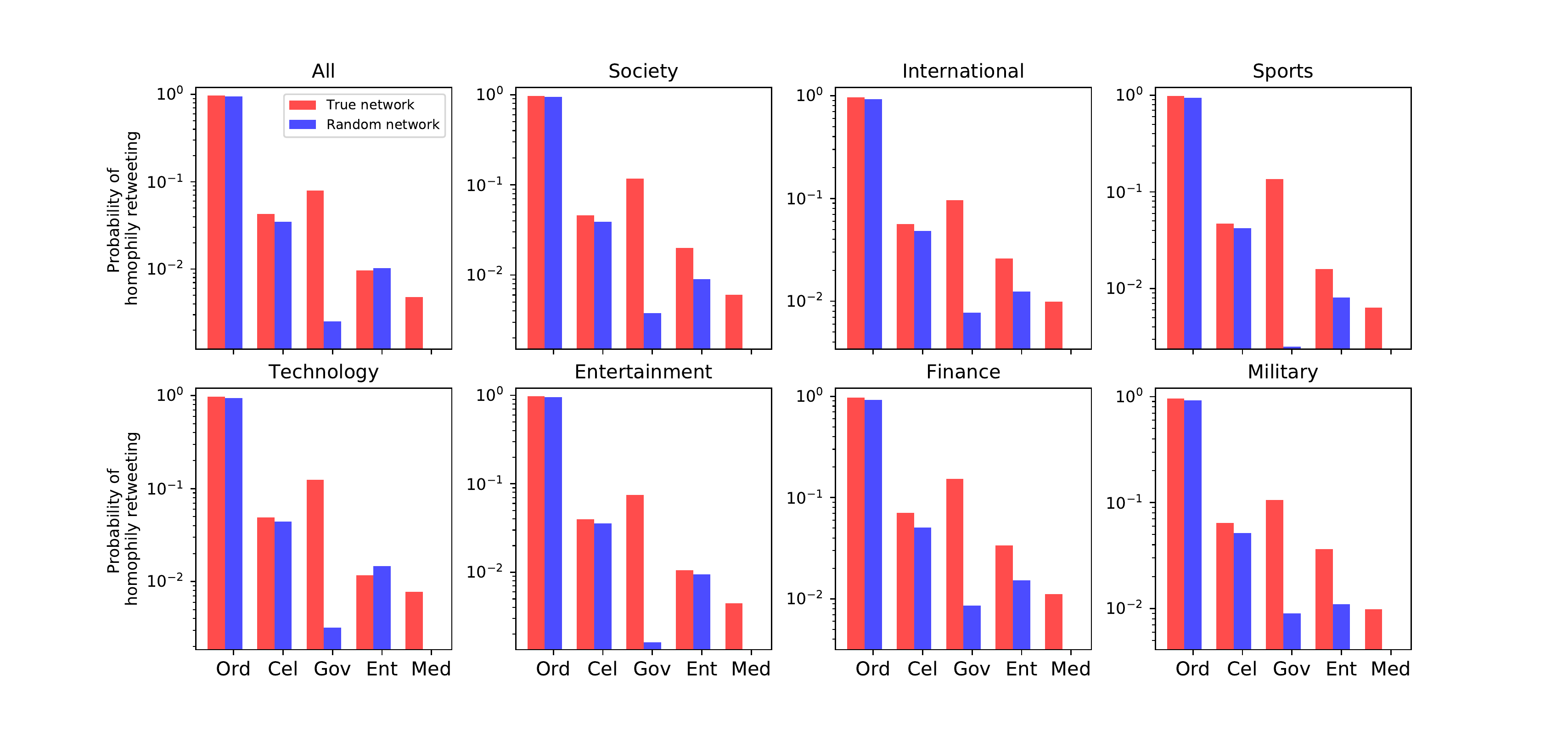}}
\quad
\subfigure[Elite]{\includegraphics[width=11cm]{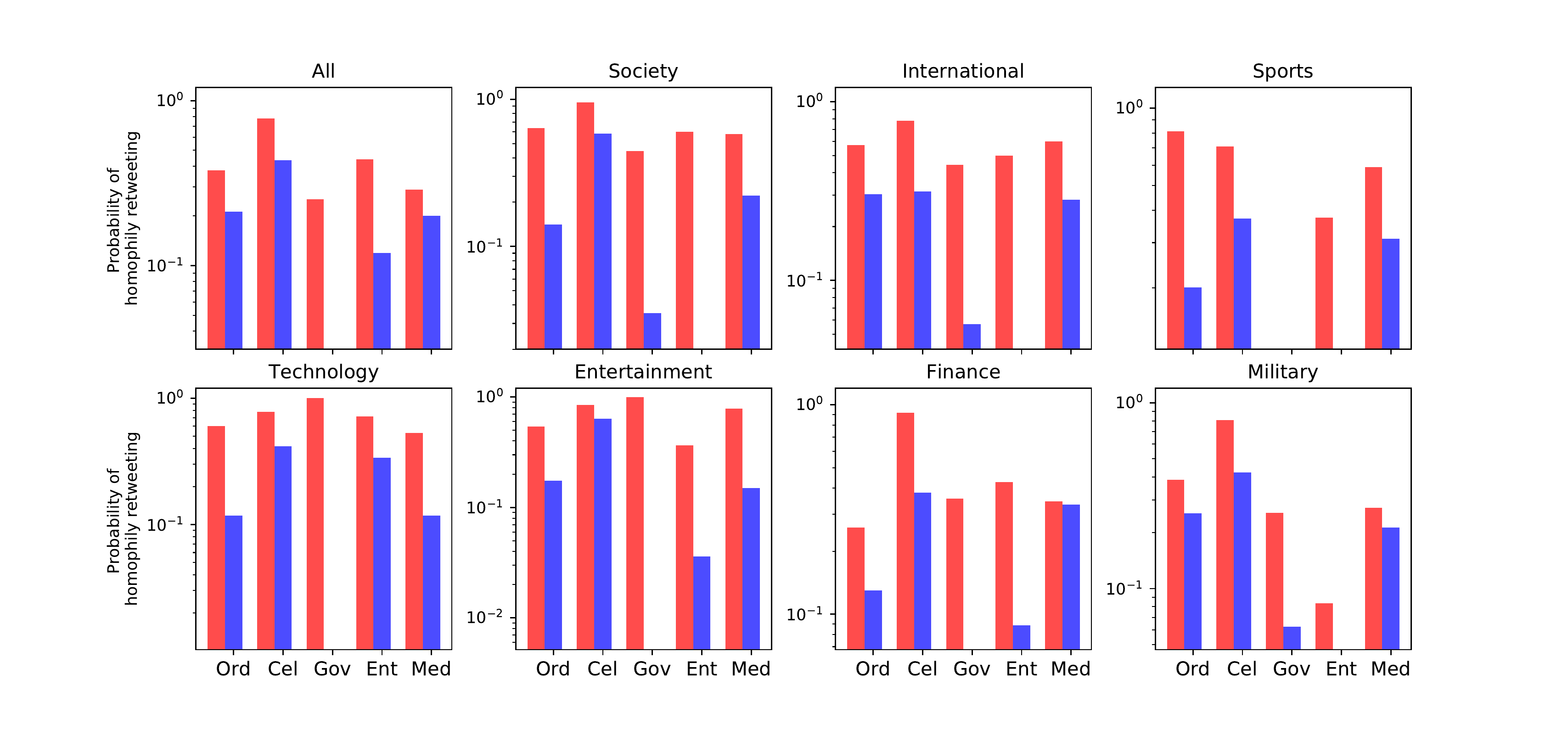}}

\caption{{\bfseries Homophily of retweeting inclination across user groups and domains.} Note that sub-graphs (a) and (b) represent the mass level and the elite level, respectively.} 
\label{fig:fig7}
\end{figure}

On the basis of the constructed forwarding network, we measure the retweeting homophily through the probability of edges connecting a pair of users with the same verified type (regardless of the edge weights). At the same time, we also calculate this indicator in the random network in which all edges are randomly rewired as a benchmark to test its significance. The comparison of homophily between the real network and its random counterpart in each domain is shown in Fig.~\ref{fig:fig7}. The homophily of government and media users is significantly higher than that in the random counterparts, indicating their inclination towards homogeneous retweeting. The group of ordinary users also possess high homophily because they account for 95\% of nodes in the network, and accordingly their random homophily is similarly high, meaning low significance. The enterprise’s homophily is even lower than the random value in the technology domain. In fact, a large part of the corporate accounts come from the emerging Internet technologies and these accounts seldom interact with other enterprises due to their competitive relationship, unless they have an interest-based partnership. Nonetheless, the homophily of elites in enterprise is always higher than the corresponding random value, which indicates that the enterprise elites can interact freely without restrictions.

\subsubsection{Loyalty}
\label{ssec:loyalty}

The weight of the edge in the network refers to the retweeting frequency which indeed reflects the loyalty of the retweeter. Considering that the tweet count of the target user will affect weights, here we use the probability that each tweet of the target user will be forwarded by each retweeter to represent the loyalty.

\begin{figure}[h!] 
\centering
\subfigure[Mass]{\includegraphics[width=11cm]{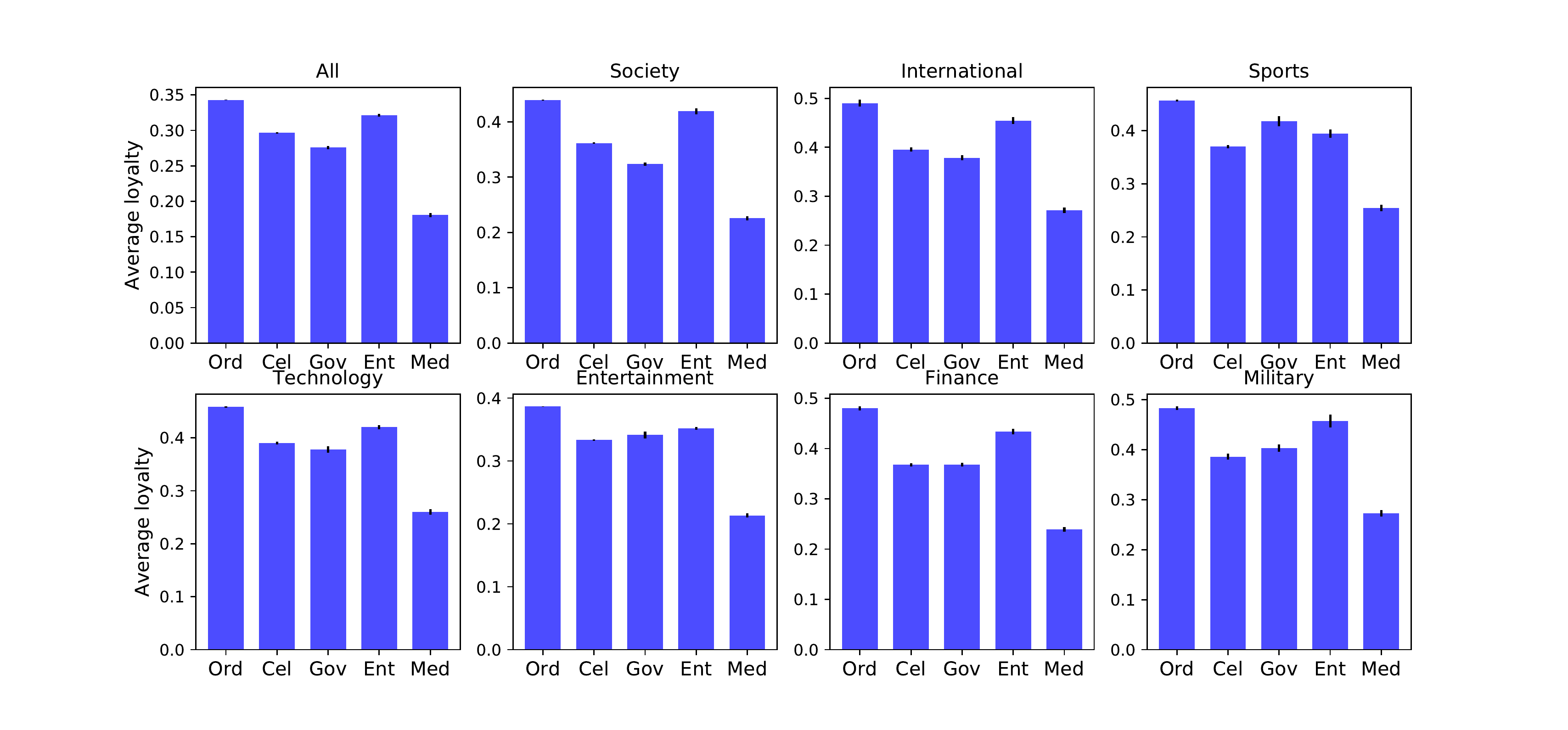}}
\quad
\subfigure[Elite]{\includegraphics[width=11cm]{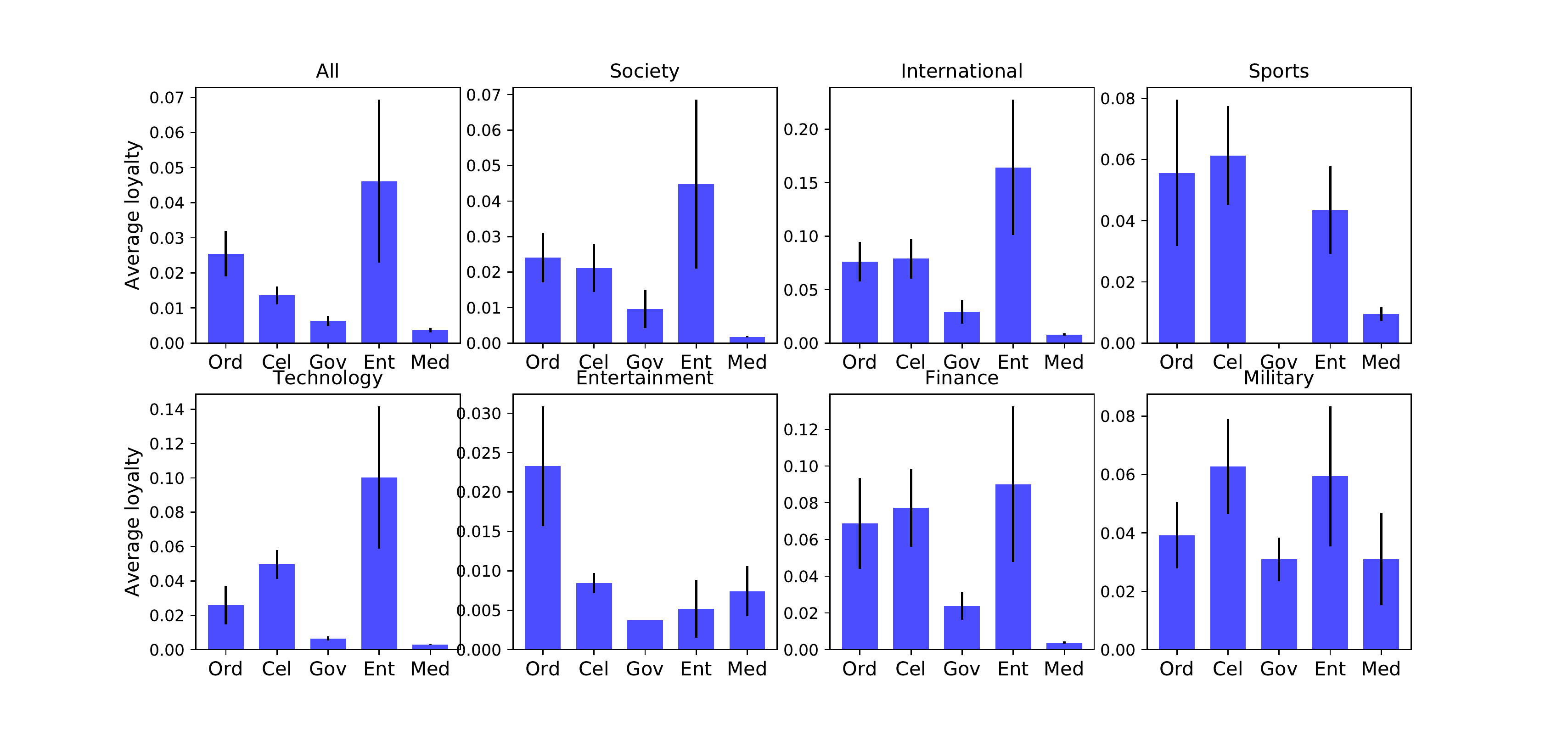}}

\caption{{\bfseries Average loyalty of retweeters across domains and user groups.} Note that sub-graphs (a) and (b) represent the mass level and the elite level, respectively.} 
\label{fig:fig8}
\end{figure}

The average loyalty of all retweeters of various user groups is shown in Fig.~\ref{fig:fig8}, where all target users have posted at least twice. It’s clear that the mass have higher average loyalty than elites, which suggests that the former are more intimate with retweeters. Interestingly, the loyalty value of media is low, which can be explained that media users attract a large number of retweeters by being active, but their audiences are less sticky. Just as a passionate fan will share almost every tweet by a star, the accounts of branch companies will keep pace with the headquarters, especially at the elite level. More importantly, the loyalty fluctuates differently at two levels, inspiring the following explorations in user-oriented promotions.

\section{User-oriented actions for popularity promotions}
\label{sec:promotions}

After the comparison of various behaviors between the mass and elites, some key actions of promoting popularity inspired by the behavioral differences are presented and testified. The experiments focus on these following questions: For users at both levels of elites and mass, what kinds of contents will obtain more retweets? How to enrich the contents? Which is more important for retweeters, the number or the loyalty?

\subsection{Content diversity}
\label{subsec:cd}

\begin{figure}[h!] 
\centering
\subfigure[Mass]{\includegraphics[width=11cm]{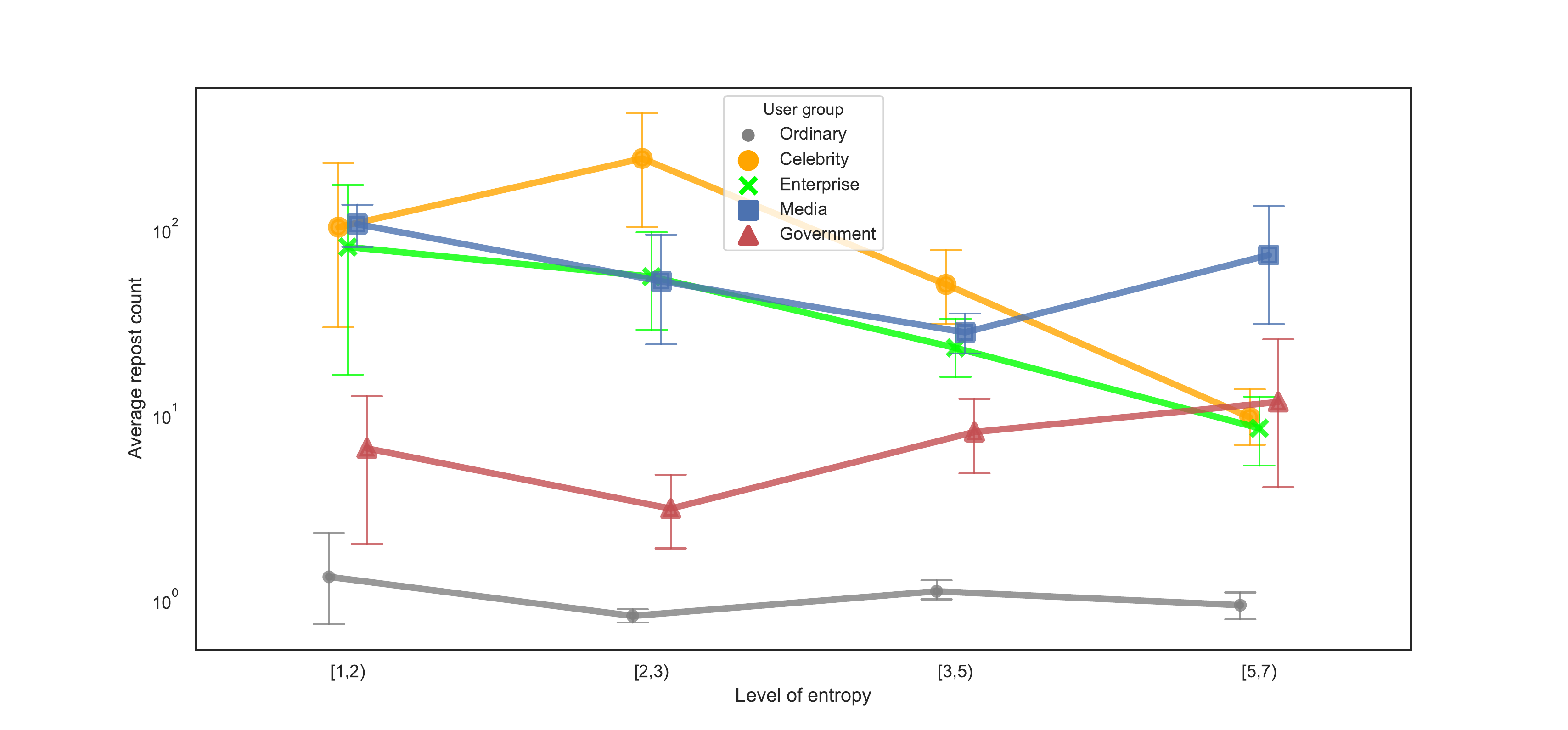}}
\quad
\subfigure[Elite]{\includegraphics[width=11cm]{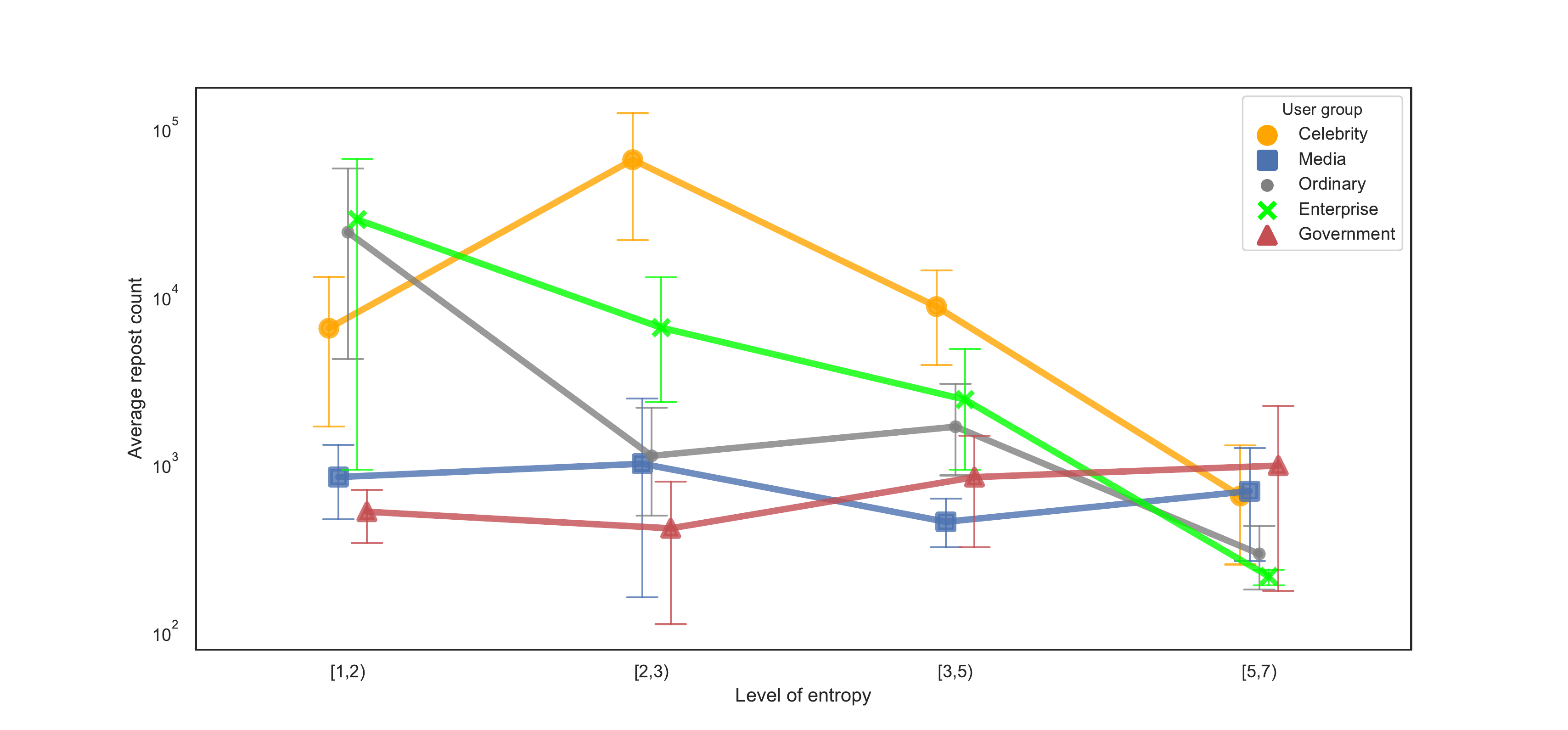}}

\caption{{\bfseries The relationship between the posting entropy and the average repost count across user groups.} Note that sub-graphs (a) and (b) represent the mass level and the elite level, respectively.} 
\label{fig:fig9}
\end{figure}

In order to explore how content diversity affects retweets, we divide the individual posting entropy into several levels. Specifically, the entropy is 0 when the user only posts in one domain, 1 represents posting two domains on average, 1.585 refers to three domains, and so on. Considering that the richer the content is, the fewer the number of users are, the grouping of users is divided into “[1, 2)”, “[2, 3)”, “[3, 5)”, and “[5, 7]” according to the corresponding entropy value of “[0, 1)”, “[1, 1.585)”, “[1.585, 2.32)”, “[2.32, 2.807]”. The average repost count and error bar of each group at the mass and elite levels are shown in Fig.~\ref{fig:fig9}, respectively. 

In general, contents with rich domains do not directly lead to more retweets, and even has a negative impact on the group of ordinary and enterprise. Enterprises are inherently professional, and ordinary users who pay attention to too many areas will be distracted \cite{zhu1992issue}, which would reduce their content quality. However, the government will slightly increase the number of retweets if they post more diverse contents, while celebrities who focus on two domains are ideal and will gain greater content popularity. Notably, the varying patterns of media and ordinary users are different at the mass and elite levels. Therefore, each group of users needs to pinpoint appropriate content domains, and can't pursue rich themes blindly.

\subsection{Content links}
\label{subsec:cl}

Upgrading or manipulating the formats of posted contents to produce “vivid” stories is another feasible path for popularity promotion. Specifically, to enhance the popularity of posted content, actions such as adding the URLs of videos, news or pictures are pervasively adopted in social media \cite{guan2014analyzing,wang2012active}. However, as we have revealed that user behavior varies across groups and domains, these actions might lose their expected effect. With the help of behavior variations across groups and domains, how to select suitable actions to enhance the popularity of contents will be illustrated in a user-oriented manner.

After merging all users’ tweets, the tweets of each user $\it{i}$ can be represented by a vector $\it{l_i=(l_{i}^{1}, l_{i}^{2}, l_i^3,l_i^4)}$, where $\it{l_i^1, l_i^2,l_i^3}$ and $\it{l_i^4}$ separately represent the fraction of tweets containing videos, news articles, pictures and non-links. For each tweet, how many times it was retweeted in our sampling period is the most convincing metric for valuing its popularity. Then, based on these preliminaries, to examine whether these enhancement actions work under circumstances of different user-domain assemblies, the pairwise Pearson correlation coefficients between $\it{l_i}$ and content popularity, i.e., the average repost count per tweet for user $\it{i}$, can be investigated at the mass and elite levels.

The results at the mass level are shown in Table 3. In neglecting content domains, the proportions containing videos are positively related to content popularity for all users except the media; this is especially the case for the government, implying a significant promotion from adding videos in tweets authored by government accounts. However, after assembling content domains and user groups into different circumstances, the effect of various actions fluctuates unexpectedly across domains. Interestingly, the enhancement effect on content popularity will be trivial in domains where users are active and be relatively significant in domains where they are inactive. For example, adding the links of videos will lead to popularity promotion for ordinary users in the technology domain, for enterprise in the military domain and for celebrity in the international domain. Meanwhile, the lack of significant results in the finance domain also suggests the possibility that these strategies might completely lose their effect under certain circumstances. Unexpectedly, actions such as adding more links of news articles might even undermine content popularity for the groups of ordinary and celebrity. This result implies the negative impact of unmatched actions in user-domain assemblies, suggesting again that behavior variations should be considered in promotion actions. 

The correlations at the elite level are presented in Table 4. In ignoring domains, adding news articles helps only enterprise, and video links of elites are not as effective as the mass, even if the former have a higher proportion of videos. However, across different user-domain assemblies, the effect of actions at the elite level demonstrates more interesting variations than those disclosed at the mass level. Specifically, on the one hand, in domains in which users are inactive, popularity will be similarly enhanced for elites. For example, adding video links will help government users earn a boost in popularity in the entertainment domain, and more links of news articles will help enterprises in the finance and military domains. On the other hand, for elites, the popularity of tweets in their active domains can also be further improved, which is inconsistent with the observations at the mass level. For example, adding the links to pictures can improve content popularity in the society domain for government users, enterprise can boost the popularity of technology-related tweets by adding links to news articles, and the financial content of media users can be popularized by adding videos. Similarly, the lack of significant results for the sports and international domains again suggests that even at the elite level, the enhancement effect of these actions might be completely lost.

Different from the conclusion reached by Wang et al. that tweets with picture links are more likely to be retweeted \cite{wang2012active}, the above results imply that these enhancement strategies actually have varying performance across user-domain assemblies. It is possible to lose the enhancement effect or to even cause negative impact if unmatched strategies are inappropriately selected. From this illustration, the variation of user behavior across domains found in this study implies that it is necessary to update previous understandings of marketing in social media. In particular, the exact mapping between behavior variations and popularity promotions will offer prior knowledge to develop appropriate strategies from a more comprehensive perspective, one in which various assemblies of user groups and content domains can practically and systematically be considered.

\subsection{Loyalty}
\label{subsec:lya}

\begin{figure}[h!] 
\centering
\subfigure[Mass]{\includegraphics[width=11cm]{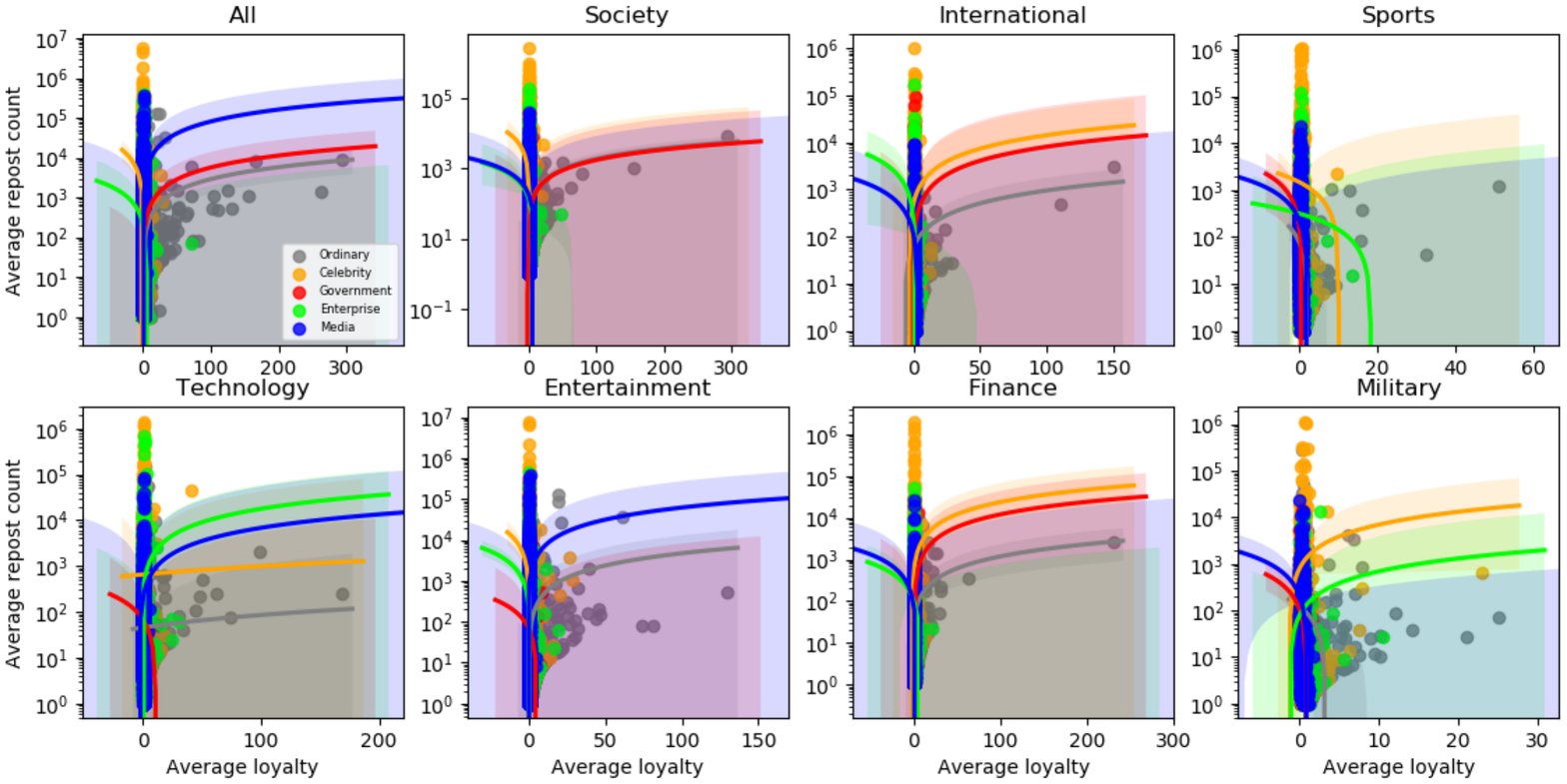}}
\quad
\subfigure[Elite]{\includegraphics[width=11cm]{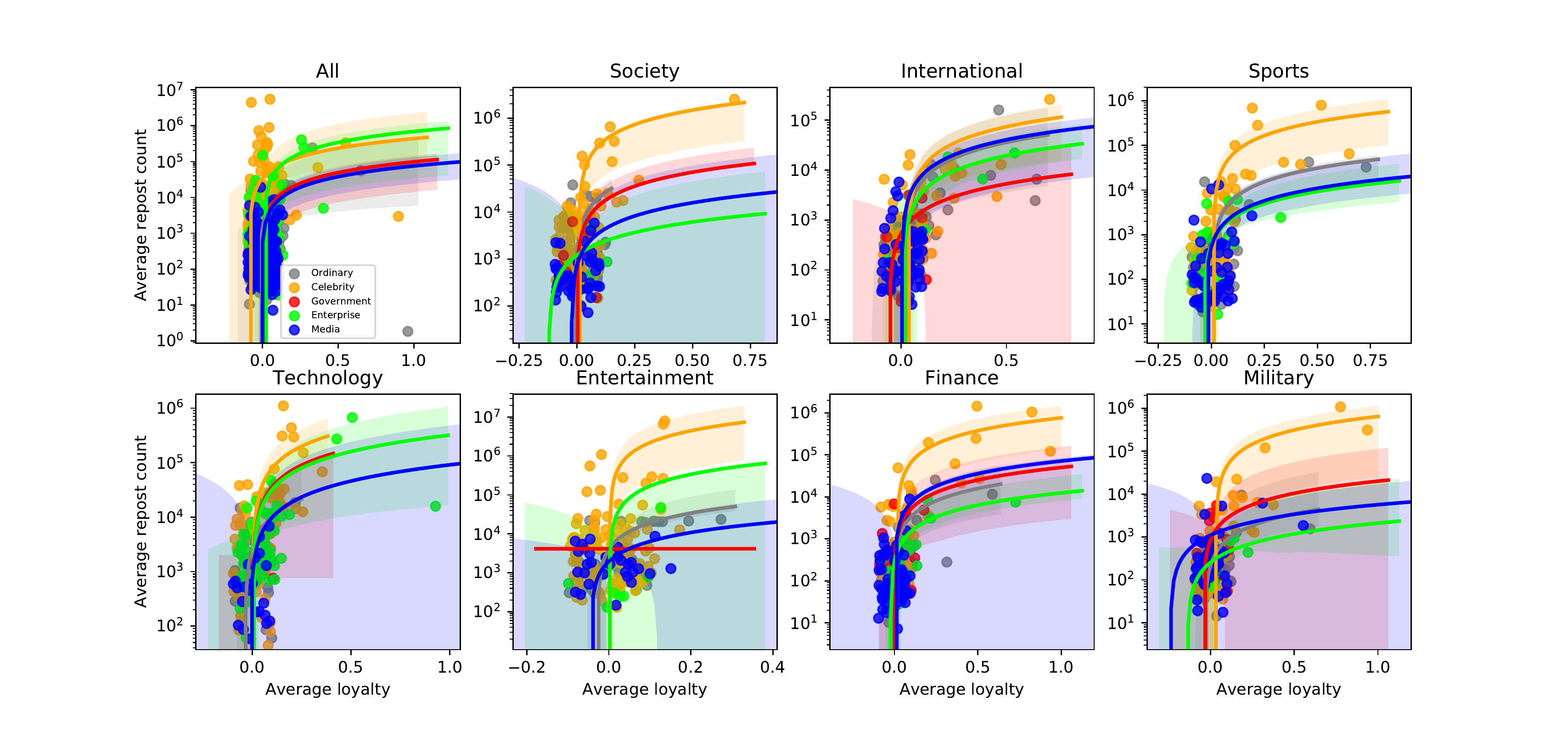}}

\caption{{\bfseries The relationship between the loyalty and the average repost count across user groups and domains.} Note that sub-graphs (a) and (b) represent the mass level and the elite level, respectively.} 
\label{fig:fig10}
\end{figure}

For each user in Weibo, the averaged retweet count per tweet is another direct reflection of the content popularity. In order to explore the impact of loyalty on the popularity of target users, the regression curves of various users at two levels are shown in Fig.~\ref{fig:fig10}. Compared with the mass, elites need more loyalty of retweeters to increase their content popularity, especially celebrities and corporates. However, the effects of loyalty on enterprise elites are unstable across domains. These patterns further indicate the heterogeneity of users even in the same verified group and suggest that the seeding of influentials and crowds is domain dependent.

Intuitively, the elites usually have a large number of retweeters which is also a key factor in popularity promotion, and this is why they were chosen. Therefore, we further explore how the retweeter count and loyalty influence the content popularity. Based on the retweet networks, a multiple regression analysis is performed with the averaged retweet count as the dependent variable, and the results of the mass and elites are shown in Table 5. From the perspective of loyalty, the coefficient of elites is higher than that of mass, implying that the loyalty of retweeter is important for elites to promote their content popularity. Moreover, the mass users should pursue more new retweeters. After all, their average loyalty is higher than elites. The results suggest that the strategies for promoting popularity of the mass and elites are significantly different from the behavior of retweeting loyalty.

Overall, from the perspectives of content diversity and links and loyalty, we establish a complete picture of the different strategies between the mass and elites across domains and groups, which can provide suitable ways for various users to increase popularity.

\section{Conclusion}
\label{sec:con}

The behavioral comparison between influential elites and the mass grassroots is an important communication issue \cite{zhang2016creates,wang2014big,su2019exploring,grossmann2018massCelite,beck2002elite}, and the popularity promotion is one of the primary goals in communication, especially in marketing scenarios \cite{figueiredo2016trendlearner,lerman2010using,rogers2010diffusion,li2018attain}. However, the behavior variations of the mass and elites across user groups and domains, and the relation between behavioral differences and strategies for enhancing popularity are rarely addressed. Meanwhile, the scarcity of massive behavioral data and expensive traditional methods make it difficult to study behavior variations in multiple domains. Fortunately, the prosperous development of social media makes it possible to collect the digital traces and interaction networks of massive users. Meanwhile, the network science and machine learning models help split 8,520,933 users into five groups, categorize over 140,000,000 tweets into seven domains, target elites with real influence and offer ideal circumstances for investigating the comprehensive mapping between behavior variations and popularity promotions. To the best of our knowledge, a complete picture of behavior variations across user groups and domains at the mass and elite levels is first established. Additionally, how diverse behaviors influence the actions for popularity promotions are thoroughly examined and testified, which can be applicable to both influentials and crowd targeting from the perspective of marketing practitioners.

There are significant differences between the mass and elites in various behavioral dimensions across user groups and domains. We find that media users are mostly active, and the varying patterns of the mass and elites are quite different across domains. In regard to temporal patterns, government, media and enterprise users have working cycles at the mass level, while the activity of elites fluctuates greatly and shows more diverse across domains than that of the mass. As for the entropy of tweeting, most elites have a wider variety of contents than the mass, and they often use video links to tell stories vividly. In addition, only the homophily of enterprises is very low. Surprisingly, the average loyalty of the mass is higher than that of elites. 

We further explored the impact of some behavioral differences on the content popularity. Unexpectedly, rich contents with domain diversity wouldn’t always bring more retweets, which is even counterproductive for enterprises and ordinary users. This result is consistent with previous study which suggests that people prefer the more professional websites in online shopping \cite{koh2010heuristic}. Moreover, correlation analysis of various links and retweets displays different communication effects across user-domain assemblies, which is not in line with the views that the picture link can increase the possibility of being retweeted \cite{wang2012active}. The promotion effect of video on the mass is stronger than that of elites who have a higher proportion of video links. Interestingly, commonly employed actions might also work well in domains in which users are inactive, which implies that shortcomings in activity can to some extent be fixed by content manipulations. For instance, government elites may gain significant popularity improvements by embedding the links of videos in their inactive domains such as entertainment and sports. Finally, we suggest that elites need to improve the quality of their fans and the mass should foster and reach new audiences to promote their popularity. 

In summary, users in social media needs to find an individualized enhancement strategy that fits their behavioral characteristics rather than mere copycat. Our findings fill the knowledge gaps of how the behavioral differences between the elite and the mass influence their marketing strategies in multiple domains, and offer guidelines on both targeting seeds and strengthening promotions in realistic marketing-like scenarios.

This paper has made a preliminary study on the relation between behavioral variations and popularity promotions, and a few limitations should be considered in reviewing the results. Although many dimensions of behavior and strategy are investigated, finding more strategies for enhance content popularity and analyzing them simultaneously is one of the future goals. In addition, the mapping discussed here is assumed to be static, and gaining an in-depth understanding of its spatiotemporal dynamics would be a promising direction for future research.

\section*{Acknowledgments}
This work was supported by NSFC (Grant Nos. 71871006 and 61421003) and the fund of the State Key Lab of Software Development Environment (Grant No. SKLSDE-2019ZX-06).

\section*{References}
\bibliographystyle{bmc-mathphys}
%\bibliography{reference}

\section*{Tables and figures}

\begin{table}[h!]
\caption{Summary statistics of user groups}
\scriptsize
\begin{tabular}{|c|c|c|c|c|c|}
\hline
User status & Ordinary  & Celebrity & Covernment & Enterprise & Media \\ \hline
\#Mass      & 8,043,807 & 301,118   & 20,370     & 87,155     & 9,983 \\ \hline
\#Elite     & 196       & 408       & 29         & 111        & 186   \\ \hline
\end{tabular}
\end{table}

\begin{table}[h!]
\caption{ANOVA of activity variations in a week. Significance levels are two-tailed; *p$<$0.05, **p$<$0.01, and ***p$<$0.001.}
\scriptsize
\resizebox{\textwidth}{!}{
\begin{tabular}{|c|c|c|c|c|c|c|c|}
\hline
Time                      & User groups & Average value of mass & Average value of elites & Sum of squares & Average square & F-value & p-value \\ \hline
\multirow{6}{*}{Weekday}  & all        & 0.719              & 0.764               & 1.011          & 1.011      & 17.487  &\bf ***     \\ \cline{2-8} 
                          & Ordinary   & 0.717              & 0.725               & 0.004          & 0.004      & 0.071   & 0.790   \\ \cline{2-8} 
                          & Celebrity  & 0.732              & 0.742               & 0.020          & 0.020      & 0.382   & 0.536   \\ \cline{2-8} 
                          & Government & 0.894              & 0.781               & 0.229          & 0.229      & 9.673   &\bf **      \\ \cline{2-8} 
                          & Enterprise & 0.831              & 0.822               & 0.005          & 0.005      & 0.110   & 0.740   \\ \cline{2-8} 
                          & Media      & 0.831              & 0.787               & 0.283          & 0.283      & 9.243   &\bf **      \\ \hline
\multirow{6}{*}{Weekends} & all        & 0.281              & 0.236               & 1.011          & 1.011      & 17.487  &\bf ***     \\ \cline{2-8} 
                          & Ordinary   & 0.283              & 0.275               & 0.004          & 0.004      & 0.071   & 0.790   \\ \cline{2-8} 
                          & Celebrity  & 0.268              & 0.258               & 0.020          & 0.020      & 0.382   & 0.536   \\ \cline{2-8} 
                          & Government & 0.106              & 0.219               & 0.229          & 0.229      & 9.673   &\bf **      \\ \cline{2-8} 
                          & Enterprise & 0.169              & 0.178               & 0.005          & 0.005      & 0.110   & 0.740   \\ \cline{2-8} 
                          & Media      & 0.169              & 0.213               & 0.283          & 0.283      & 9.243   &\bf **      \\ \hline
\end{tabular}}
\end{table}

\begin{table}[]
\caption{Pearson correlation coefficients between actions and popularity at the mass level. User groups and content domains are assembled to simulate various circumstances. Significance levels are two-tailed; *p$<$0.05, **p$<$0.01, and ***p$<$0.001.}
\scriptsize
\begin{tabular}{|c|c|c|c|c|}
\hline
Content domains                & All users  & Video    & News article & Picture  \\ \hline
\multirow{5}{*}{All}           & Ordinary   &\bf 0.013*** & -0.006       & 0.002    \\ \cline{2-5} 
                               & Celebrity  &\bf 0.050*** & -0.024       & -0.006   \\ \cline{2-5} 
                               & Government &\bf 0.224**  & -0.043       & 0.114    \\ \cline{2-5} 
                               & Enterprise &\bf 0.096**  & -0.043       & 0.038    \\ \cline{2-5} 
                               & Media      & 0.072    & -0.065       & 0.018    \\ \hline
\multirow{5}{*}{Society}       & Ordinary   & 0.003    & -0.006       & 0        \\ \cline{2-5} 
                               & Celebrity  &\bf 0.123*** & -0.025       & -0.005   \\ \cline{2-5} 
                               & Government & 0.071    & -0.059       & 0        \\ \cline{2-5} 
                               & Enterprise & -0.01    & 0.071        & 0        \\ \cline{2-5} 
                               & Media      & -0.038   & 0.073        & 0        \\ \hline
\multirow{5}{*}{International} & Ordinary   & 0.006    & 0            & 0        \\ \cline{2-5} 
                               & Celebrity  &\bf 0.080**  & 0.015        & -0.005   \\ \cline{2-5} 
                               & Government & 0.025    & 0.054        & 0.056    \\ \cline{2-5} 
                               & Enterprise & -0.012   & 0.084        & -0.013   \\ \cline{2-5} 
                               & Media      & 0.164    & -0.073       & -0.003   \\ \hline
\multirow{5}{*}{Sports}        & Ordinary   &\bf 0.041*** &\bf -0.019***    & 0.001    \\ \cline{2-5} 
                               & Celebrity  & 0.04     & -0.021       & -0.005   \\ \cline{2-5} 
                               & Government & 0.062    & 0.036        & 0        \\ \cline{2-5} 
                               & Enterprise & 0.089    & -0.061       & 0.004    \\ \cline{2-5} 
                               & Media      & -0.046   & -0.059       &\bf 0.489*** \\ \hline
\multirow{5}{*}{Technology}    & Ordinary   &\bf 0.017*** &\bf -0.020***    & 0.004    \\ \cline{2-5} 
                               & Celebrity  & 0.002    & -0.024       & -0.003   \\ \cline{2-5} 
                               & Government & 0.149    & 0.002        & 0.08     \\ \cline{2-5} 
                               & Enterprise & 0.017    & -0.039       & -0.005   \\ \cline{2-5} 
                               & Media      & -0.011   & 0.02         & 0.01     \\ \hline
\multirow{5}{*}{Entertainment} & Ordinary   &\bf 0.010**  &\bf -0.009*      & 0.003    \\ \cline{2-5} 
                               & Celebrity  & 0.03     &\bf -0.038*      & -0.008   \\ \cline{2-5} 
                               & Government & 0.057    & -0.062       & -0.025   \\ \cline{2-5} 
                               & Enterprise & -0.001   & -0.04        & -0.003   \\ \cline{2-5} 
                               & Media      & 0.05     & -0.061       & 0.008    \\ \hline
\multirow{5}{*}{Finance}       & Ordinary   & 0.005    & -0.006       & 0        \\ \cline{2-5} 
                               & Celebrity  & 0.028    & -0.022       & -0.004   \\ \cline{2-5} 
                               & Government & 0.01     & -0.08        & 0.034    \\ \cline{2-5} 
                               & Enterprise & -0.008   & -0.012       & -0.005   \\ \cline{2-5} 
                               & Media      & -0.051   & -0.028       & -0.052   \\ \hline
\multirow{5}{*}{Military}      & Ordinary   & 0.005    &\bf -0.019**     & 0.003    \\ \cline{2-5} 
                               & Celebrity  & 0.006    & -0.02        & -0.008   \\ \cline{2-5} 
                               & Government & 0.145    & -0.018       & -0.04    \\ \cline{2-5} 
                               & Enterprise &\bf 0.316*** & 0.008        & 0        \\ \cline{2-5} 
                               & Media      & -0.043   & -0.12        & -0.024   \\ \hline
\end{tabular}
\end{table}

\begin{table}[]
\caption{Pearson correlation coefficients between actions and popularity at the elite level. User groups and content domains are assembled to simulate various circumstances. Significance levels are two-tailed; *p$<$0.05, **p$<$0.01, and ***p$<$0.001.}
\scriptsize
\begin{tabular}{|c|c|c|c|c|}
\hline
Content domains                & Elites     & Video   & News article & Picture  \\ \hline
\multirow{5}{*}{All}           & Ordinary   & -0.075  & -0.081       & -0.019   \\ \cline{2-5} 
                               & Celebrity  & -0.073  & 0.004        & -0.023   \\ \cline{2-5} 
                               & Government & -0.003  & 0.16         & 0.272    \\ \cline{2-5} 
                               & Enterprise & -0.142  &\bf 0.240*       & -0.032   \\ \cline{2-5} 
                               & Media      & 0.124   & -0.085       & -0.042   \\ \hline
\multirow{5}{*}{Society}       & Ordinary   & -0.058  & -0.049       & -0.012   \\ \cline{2-5} 
                               & Celebrity  & -0.094  & 0.096        & -0.008   \\ \cline{2-5} 
                               & Government & -0.166  & 0.21         &\bf 0.674*** \\ \cline{2-5} 
                               & Enterprise & 0.188   & -0.003       & -0.028   \\ \cline{2-5} 
                               & Media      & 0.078   & -0.096       & -0.032   \\ \hline
\multirow{5}{*}{International} & Ordinary   & -0.014  & -0.042       & -0.02    \\ \cline{2-5} 
                               & Celebrity  & -0.061  & -0.055       & -0.008   \\ \cline{2-5} 
                               & Government & 0.263   & -0.197       & 0.013    \\ \cline{2-5} 
                               & Enterprise & 0.034   & -0.061       & 0        \\ \cline{2-5} 
                               & Media      & 0.137   & -0.058       & -0.043   \\ \hline
\multirow{5}{*}{Sports}        & Ordinary   & -0.139  & -0.082       & -0.033   \\ \cline{2-5} 
                               & Celebrity  & -0.072  & -0.041       & -0.012   \\ \cline{2-5} 
                               & Government & 0.103   & -0.02        & 0        \\ \cline{2-5} 
                               & Enterprise & -0.088  & 0.182        & -0.001   \\ \cline{2-5} 
                               & Media      & 0.041   & -0.025       & -0.037   \\ \hline
\multirow{5}{*}{Technology}    & Ordinary   & -0.1    &\bf -0.148*      & -0.015   \\ \cline{2-5} 
                               & Celebrity  & -0.031  & 0.018        & -0.016   \\ \cline{2-5} 
                               & Government & 0.048   & 0.s147        & -0.113   \\ \cline{2-5} 
                               & Enterprise & -0.089  &\bf 0.302**      & -0.022   \\ \cline{2-5} 
                               & Media      &\bf 0.150*  & -0.074       & -0.029   \\ \hline
\multirow{5}{*}{Entertainment} & Ordinary   & -0.048  &\bf -0.189**     & -0.028   \\ \cline{2-5} 
                               & Celebrity  & -0.065  & 0.005        & -0.013   \\ \cline{2-5} 
                               & Government &\bf 0.502** & -0.024       & 0.006    \\ \cline{2-5} 
                               & Enterprise & -0.105  & -0.028       & -0.028   \\ \cline{2-5} 
                               & Media      & 0.07    & -0.087       & -0.033   \\ \hline
\multirow{5}{*}{Finance}       & Ordinary   & -0.045  & -0.055       & -0.012   \\ \cline{2-5} 
                               & Celebrity  & -0.031  & -0.058       & -0.007   \\ \cline{2-5} 
                               & Government & -0.057  & -0.075       & -0.09    \\ \cline{2-5} 
                               & Enterprise & 0.125   &\bf 0.358**      & 0        \\ \cline{2-5} 
                               & Media      &\bf 0.165*  & 0.019        & -0.007   \\ \hline
\multirow{5}{*}{Military}      & Ordinary   & 0.095   & -0.069       & -0.041   \\ \cline{2-5} 
                               & Celebrity  & -0.071  & -0.045       & -0.011   \\ \cline{2-5} 
                               & Government & -0.056  & -0.031       & 0.334    \\ \cline{2-5} 
                               & Enterprise & -0.077  &\bf 0.324**      & -0.025   \\ \cline{2-5} 
                               & Media      & 0.057   & 0            & -0.018   \\ \hline
\end{tabular}
\end{table}

\begin{table}[ht!]
\caption{Results of multiple regression analysis at the mass and elite levels. Significance levels are two-tailed; *p$<$0.05, **p$<$0.01, and ***p$<$0.001.}
\scriptsize
\begin{tabular}{|l|l|l|l|l|l|l|l|l|}
\hline
                & \multicolumn{4}{c|}{Mass}                                                                                                   & \multicolumn{4}{c|}{Elite}                                                                                                  \\ \hline
                & \multicolumn{1}{c|}{coef} & \multicolumn{1}{c|}{std err} & \multicolumn{1}{c|}{$t$} & \multicolumn{1}{c|}{p\textgreater{}$\left|t\right|$} & \multicolumn{1}{c|}{coef} & \multicolumn{1}{c|}{std err} & \multicolumn{1}{c|}{$t$} & \multicolumn{1}{c|}{p\textgreater{}$\left|t\right|$} \\ \hline
const           & 1.2787                    & 0.007                        & 193.957                &\bf ***                                     & 22.8254                   & 10.615                       & 2.15                   & (0.032)*                                \\ \hline
average loyalty & 0.8224                    & 0.012                        & 66.248                 &\bf ***                                     & 1629.9279                 & 90.067                       & 18.097                 &\bf ***                                     \\ \hline
retweeter count & 0.0037                    & 0                            & 340.457                &\bf ***                                     & 0.0018                    & 0                            & 4.291                  &\bf ***                                     \\ \hline
observations    & \multicolumn{4}{l|}{3024960}                                                                                                & \multicolumn{4}{l|}{928}                                                                                                    \\ \hline
$R^2$              & \multicolumn{4}{l|}{0.038}                                                                                                  & \multicolumn{4}{l|}{0.267}                                                                                                  \\ \hline
adjust $R^2$       & \multicolumn{4}{l|}{0.038}                                                                                                  & \multicolumn{4}{l|}{0.266}                                                                                                  \\ \hline
F-statistic     & \multicolumn{4}{l|}{59660}                                                                                                  & \multicolumn{4}{l|}{168.6}                                                                                                  \\ \hline
\end{tabular}
\end{table}

\end{document}